\documentclass{article}
\usepackage{amssymb}
\usepackage{amsmath}
\usepackage{graphicx}
\usepackage{amsmath,amssymb}

\input{tcilatex}
\begin{document}

\title{Laser Induced Heat Diffusion Limited Tissue Coagulation:\\
Problem and General Properties}

\author{I. A. Lubashevsky \and V. V. Gafiychuk  \and A. V. Priezzhev  \\
Institute for Applied Problems in Mechanics and Mathematics,\\
National Academy of Sciences of Ukraine,\\
3b Naukova str. Lviv, 290601, Ukraine.\\
Theory Department, General Physics Institute,\\
Russian Academy of Sciences, \\
Vavilov str. 38, Moscow 117942, Russia.\\
Physics Department, Moscow State University,\\
Vorobievy Gory, Moscow 119899, Russia.
}

\date{}
\maketitle

\begin{abstract}
Previously we have developed a free boundary model for local thermal
coagulation induced by laser light absorption when the tissue region
affected directly by laser light is sufficiently small and heat diffusion
into the surrounding tissue governs the necrosis growth. In the present
paper surveying the obtained results we state the point of view on the
necrosis formation under these conditions as the basis of an individual
laser therapy mode exhibiting specific properties. In particular, roughly
speaking, the size of the resulting necrosis domain is determined by the
physical characteristics of the tissue and its response to local heating,
and by the applicator form rather than the treatment duration and the
irradiation power.
\end{abstract}


\section{What the heat diffusion limited tissue coagulation is: a phenomenon
and a theoretical problem}

Thermal coagulation of living tissue caused by local heating due to laser
light absorption is one of the novel thermotherapy techniques of tumor
treatment which is currently undergoing clinical trials (see, e.g., \cite{I3}%
). Thermal coagulation is used to form a necrosis domain of desired form for
the removal of the malignant tissue. So the mathematical modeling of the
necrosis growth is required, first, to find out the physical limitations and
the basic characteristics of the treatment and, second, to optimize the
therapy course. However, living tissue is extremely complex in structure,
thereby, for the adequate theoretical model to be developed and for the
mathematical modeling of the given process to be implemented reliably
typical limit cases should be singled out and studied individually. Such a
case is the subject of the present paper.

\begin{figure}[tbp]
\par
\begin{center}
\includegraphics[width=80mm]{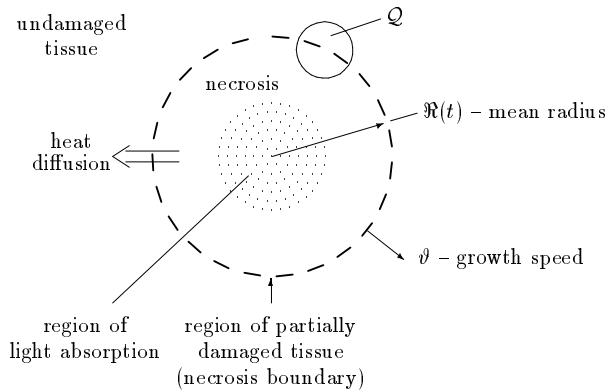}
\end{center}
\caption{The necrosis growth due to local thermal tissue coagulation limited
by heat diffusion in the surrounding undamaged tissue.}
\label{ThC:F1}
\end{figure}

When the tissue region affected directly by laser light is sufficiently
small is size, the necrosis formation is mainly governed by heat diffusion.
Namely, in this case we actually deal with the following physical process
(Fig.~\ref{ThC:F1}). Absorption of laser light delivered into a small
internal region of living tissue causes the temperature to attain such high
values (about or above $70~^{\circ}$C) that lead practically to immediate
thermal coagulation in this region. Heat diffusion into the surrounding live
tissue causes its further coagulation, giving rise to the growth of the
necrosis domain. In this case heat diffusion plays a significant role in the
necrosis growth because the necrosis size $\Re $ exceeds the depth of laser
light penetration into the tissue. Therefore, the temperature distribution
inevitably has to be substantially nonuniform and for the tissue to
coagulate at peripheral points heat diffusion should cause the temperature
to grow at these points. The latter property singles out the specific mode
of thermal coagulation under discussion from other possible types of
thermotherapy treatment and that is why we refer to the necrosis growth
under the given conditions as to thermal tissue coagulation limited by heat
diffusion. In particular, as will be seen below, the optimal implementation
of this thermotherapy mode is characterized by the formation of necrosis
domains of size $\Re \sim 5${--}$10$~mm and by the treatment duration of $t_{%
\text{course}}\sim \Re ^{2}/D\sim 2${--}$8$~min (where $D\approx 2\cdot
10^{-3}${\thinspace cm}$^{2}$/sec is the tissue temperature diffusivity).
Laser lighted is considered to be absorbed within a layer of thickness less
than $\Re$.

Heat diffusion in the live tissue is affected substantially by blood
perfusion causing the heat sink \cite{P48,CH80}. Thus, the characteristics
of the spatial distribution and the dynamics of the blood perfusion rate
should have a substantial effect on the necrosis growth limited by heat
diffusion. Therefore, in modeling mode one has to take into account the
tissue response to the temperature growth which can locally give rise to a
tenfold increase in the blood perfusion rate \cite{Song84}. The latter
effect, in particular, is responsible for a substantially nonuniform
distribution of the blood perfusion rate and visually manifests itself in a
red ring (``hyperemic ring'') appearing around the necrosis region during
the thermotherapy treatment. Besides, when heated the living tissue will
inevitably exhibit spatial nonuniformities in the temperature due to the
vessel discreteness \cite{Bi86}. Because of the extremely strong temperature
dependence of the thermal coagulation rate such temperature nonuniformities
can substantially perturb the necrosis form and, so, this effect should be
also taken into account.

In the previous papers \cite{we1,we2,we3,we4,we5,we6,we7,we8,we9} basing on
the free boundary description we have developed a model for the heat
diffusion limited thermal tissue coagulation and studied the properties of
the corresponding necrosis growth. This model has been developed in order to
take into account the effects mentioned above. The aim of the present paper
is to survey the obtained results, to outline the general properties of the
necrosis growth under the aforementioned conditions, and to justify the
point of view on the laser induced heat diffusion limited tissue coagulation
as a individual mode of laser therapy.

However, before stating the free boundary model let us, first, recall the
main features of living tissue affecting heat transfer, discuss the approach
previously used by other authors and analyze the corresponding problems in
the theoretical description of the local thermal coagulation. In this way we
will make the key points of the theory we have developed more clear.

\subsection{Background: the main features of bioheat transfer\label%
{ThC:sec.B}}

\subsubsection{Living tissue as a heterogeneous medium\label{ThC:subsec.HM}}

Blood flowing through vessels in living tissue forms paths of fast heat
transport and under typical conditions it is blood flow that governs heat
propagation on scales exceeding several millimeters (for an introduction to
this problem see, e.g., \cite{CH80,SE70}). Blood vessels make up a complex
network being practically a fractal. The larger is a vessel, the faster is
the blood motion in it and, so, the stronger is the effect of blood flux in
the given vessel on heat transfer. The blood flux in the smallest vessels,
capillaries, practically does not affect heat propagation \cite{CH80}. Thus,
there should be blood vessels of a certain length $\ell _{v}$ that are the
smallest among the vessels wherein the blood flux affects heat transfer
remarkably. The value of $\ell _{v}$ may be different at various points
(under nonuniform blood perfusion) and can be found from the expression \cite
{BOOK} (see also \cite{CH80,SE70}):
\begin{equation}
\ell _{v}(\mathbf{r})\approx \sqrt{\frac{\kappa }{c\rho f\,j_{v} (\mathbf{r}%
)L_{n}}}\,.  \label{ThC:e1.4}
\end{equation}
Here $\kappa $ is the thermal conductivity of the cellular tissue, $c$, $%
\rho $ are the density and heat capacity of the tissue, $j$ is the blood
perfusion rate (the volume of blood going through tissue region of unit
volume per unit time) and $j_{v}$ is its value averaged over scales of order
of $\ell _{v}(\mathbf{r})$. The factor $f<1$ accounts for the
counter-current effect. Initially it was phenomenologically introduced in
the bioheat equation as a certain renormalization of the blood perfusion
rate \cite{C80,W87a,W87b} and its theoretical estimate will be discussed
below in the present section. The factor $L_{n}$ playing a key role in the
theory of bioheat transfer is given by the formula:
\begin{equation}
L_{n}=\ln \Bigl(\frac{l}{a}\Bigr)\,,  \label{ThC:e1.5}
\end{equation}
where $l/a$ is the mean ratio of the individual length to radius of blood
vessels forming peripheral systems of blood circulation. Since expression~(%
\ref{ThC:e1.4}) estimating the length $\ell_v$ contains the blood perfusion
rate $j_v$ itself averaged over the scale $\ell_v$, it is implicit and has
to be completed with an expression relating the value $\ell_v(\mathbf{r})$,
the averaged perfusion rate $j_v(\mathbf{r})$ and the true one $j(\mathbf{r}%
) $. At first, we write it in a symbolic form
\begin{equation}
\mathcal{\hat{L}}\left\{ \ell _{v},j_{v},j\right\} =0\,.  \label{ThC:e1.6}
\end{equation}
Nevertheless, to estimate the length $\ell_v$ we may ignore the difference
between the averaged and true blood perfusion rates in expression~(\ref
{ThC:e1.4}). Then for the typical values of the ratio $l/a\sim 20$--40 \cite
{Mch89}, the thermal conductivity $\kappa \sim 7\cdot 10^{-3}$ \thinspace
W/cm$\cdot$K, the heat capacity $c\sim 3.5$\thinspace J/g$\cdot $ K, and the
density $\rho \sim 1$\thinspace g/cm$^{3}$ of the tissue, as well as setting
the blood perfusion rate $j_{v}\sim j\sim 0.3$\thinspace min$^{-1}$ and the
factor $f\sim 0.5$ we get from (\ref{ThC:e1.4}) and (\ref{ThC:e1.5}):
\begin{equation}
\ell _{v}\sim 4\,\mathrm{mm}\quad \mathrm{and}\quad L_{n}\approx 3-4\,.
\label{ThC:e1.7}
\end{equation}

The theory of heat transfer in living tissue (bioheat transfer theory)
starts from the system of microscopic equations governing heat propagation
in the cellular tissue and with blood inside the vessels individually.
However, on one hand, the complex structure of the vascular network and, on
the other hand, the lack of the detailed information about the arrangement
of individual vessels necessitate the development of the macroscopic
description of heat transfer in living tissue regarded as an effective
continuum with certain, may be, anomalous properties. Moreover, the vascular
network can vary in particular details from tissue to tissue or even from
patient to patient for one tissue. In this case the development of an
adequate macroscopic theory is the only way to model a thermotherapy
treatment. The macroscopic governing equation for the tissue temperature can
be obtained by averaging the microscopic equations over certain scales that,
first, are small enough so the tissue temperature does not exhibit great
variation on these scales. Second, these scales should be large enough for
the heat exchange between the cellular tissue and blood could be treated
within a continuous approximation. This heat exchange is directly controlled
by the vessels of length $\ell _{v}$. So the length $\ell _{v}$ is a natural
scale of the averaging in the theory of bioheat transfer. Besides, it turns
out that the characteristic scale of the temperature variations is
\begin{equation}
\ell _{T}\sim \sqrt{L_{n}}\ell _{v}\sim \sqrt{\frac{\kappa }{c\rho f\,j}}\,.
\label{ThC:e1.7a}
\end{equation}
The latter estimate demonstrates us that the value $1/L_{n}$ plays the role
of small parameter in the bioheat transfer theory, exactly which has made it
feasible to construct a mean field approximation regarding living tissue as
a continuum \cite{BOOK}.

In particular, averaging the microscopic equations in this way, we have got
the following macroscopic bioheat equation relating the tissue temperature $%
T_{v}$ averaged over scales of order of $\ell _{v}$ and the averaged
perfusion rate $j_{v}$:
\begin{equation}
c\rho \frac{\partial T_{v}}{\partial t}=\nabla \left( \kappa _{\text{eff}
}\nabla T_{v}\right) -fc\rho j_{v}(T_{v}-T_{a})+q_{v}\,,  \label{ThC:e1.8}
\end{equation}
where we have ignored the difference in the density and heat capacity of the
tissue and blood, respectively, $c\simeq c_{b}$, $\rho \simeq \rho _{b}$,
the blood temperature $T_a$ in large arteries of systemic circulation is
considered to be fixed, and $q_{v}$ is the heat generation rate averaged
over the same scales. The factor $f$, accounting for the counter-current
effect, and the ratio $F=\kappa _{\text{eff}}/\kappa $ are estimated by the
expressions:
\begin{equation}
f\sim \frac{1}{\sqrt{L_{n}}}\sim 0.5\,,\quad F=F(L_{n})\sim 2\,.
\label{ThC:e1.9}
\end{equation}
It should be pointed out that the factors $f$ and $F$ are determined solely
by the geometry of the vascular network, which directly results from the
vascular network being fractal in structure \footnote{%
The theoretical estimate of $f\sim 0.5$ has been obtained for the first time
in book \cite{BOOK}, announced in our papers \cite{we1,we3} and has been
later presented independently in paper \cite{WXZE97}. The effect of blood
perfusion on the heat propagation rate in terms of the heat conductivity
renormalization was quantitatively studied actually for the first time in
paper \cite{WJ85}}. Moreover, due to the logarithmically weak dependence~(%
\ref{ThC:e1.5}) these factors can be approximately treated as constants
determined beforehand.

It should be noted that blood flow through the large arteries can give rise
to an anomalously fast heat transport over scales much greater than $\ell
_{T}$. Under certain conditions it may appear that the effective thermal
conductivity $\kappa_{\mathrm{eff}}$ exceeds the thermal conductivity of the
cellular tissue by tenfold due to this effect. However, such a fast heat
transport cannot be described in terms of the mean field theory and deserves
an individual consideration. Besides, on the average, its role is not too
essential because of the sufficiently small relative volume of the large
arteries. Therefore dealing with local thermal coagulation we can ignore
this effect at the first approximation. We also point out that equation~(\ref
{ThC:e1.8}) is similar in its form to the phenomenological generalized
bioheat equation \cite{L90} except for the latter contains the true
perfusion rate $j$ rather than the averaged one $j_{v}$. This fact can be
essential for substantially nonuniform blood perfusion \cite{BOOK} as it is
the case in local thermal coagulation \cite{we1,we3}.

The mean field approximation leading to equation~(\ref{ThC:e1.8}) does not
allow for the temperature nonuniformities caused by the vessel discreteness
\cite{Bi86} because it regards living tissue as a certain continuum. So, to
take into account these nonuniformities we have to go beyond the scope of
the mean field theory. Since the particular details of the vessel
arrangement on scales about several millimeters are practically unknown and,
moreover, alter in various tissues and, may be, at different points of one
tissue it is reasonable to regard the vessel arrangement and the
corresponding temperature nonuniformities as random \cite{Bi94}. In
particular, under uniform heating, i.e. when $T_{v}$ is constant, these
nonuniformities will be characterized by the mean amplitude $\sigma $ and
the correlation length $\lambda $:
\begin{eqnarray}  \label{ThC:e1.11}
\left\langle \delta T(\mathbf{r},t)\delta T(\mathbf{r},t)\right\rangle &=&
\sigma ^{2}\quad \mathrm{and} \\
\left\langle \delta T(\mathbf{r} ,t)\delta T(\mathbf{r}^{\prime
},t)\right\rangle &\ll& \sigma ^{2}\quad \mathrm{for}\quad \left| \mathbf{r}-%
\mathbf{r}^{\prime }\right| \gg \lambda \,,  \notag
\end{eqnarray}
where the symbol $\left\langle \ldots \right\rangle $ stands for averaging.
As might be expected, the correlation length $\lambda $ is about $\ell _{v}$
($\lambda \sim \ell _{v}$) and depends on the local value of the perfusion
rate. By contrast, due to the fractal structure of the vascular network the
ratio of the amplitude $\sigma $ to the overheating $T_{v}-T_{a}$ is
estimated as \cite{BOOK}:
\begin{equation}
\frac{\sigma }{T_{v}-T_{a}}\lesssim \frac{1}{L_{n}}  \label{ThC:e1.10}
\end{equation}
and can be regarded as a predetermined constant. Taking into account
additional numerical factors \cite{BOOK} we get $\sigma \sim (0.1$--$%
0.2)(T-T_{a})$ (see also \cite{Bi86}). Since the vessel discreteness effect
is mainly caused by the vessels of lengths about $\ell _{v}$ it can be
described in terms of a bioheat equation similar to (\ref{ThC:e1.8}) which,
however, contains the true tissue temperature $T$ and where the averaged
perfusion rate $j_{v}$ is replaced by term:
\begin{equation}
j_{v}\rightarrow j_{v}+\delta j(\mathbf{r})\,,  \label{ThC:e1.12}
\end{equation}
involving the random component $\delta j(\mathbf{r})$ meeting the
conditions:
\begin{eqnarray}
\left\langle \delta j(\mathbf{r})\right\rangle &=&0\,,  \label{ThC:e1.13a} \\
\left\langle \delta j(\mathbf{r})\delta j(\mathbf{r}^{\prime })\right\rangle
&=&j_{v}^{2}g^{0}\left( \frac{\left| \mathbf{r}-\mathbf{r}^{\prime }\right|
}{\ell _{v}}\right) \,.  \label{ThC:e1.13b}
\end{eqnarray}
Hear the function $g^{0}(x)$ is such that $\left|g^{0}(x)\right|\sim 1$ for $%
x\lesssim 1$ and $g^{0}(x)\ll 1$ for $x\gg 1$.

In this subsection we have discussed the characteristic features of living
tissue as a heterogeneous medium. However it is also an active medium
because it responds to temperature variations by increasing the blood vessel
radius, which gives rise to an increase in the blood perfusion rate $j$. The
blood perfusion rate can locally grow by tenfold \cite{Song84}, so, this
effect is significant and is the subject of the next subsection.

\subsubsection{Living tissue as an active medium\label{ThC:subsec.AM}}

Living tissue tries to keep its temperature from exceeding a certain vital
boundary $T_{+}\approx 44$--$46$~$^{\circ}$C ($T<T_{+}$) to prevent thermal
damage (see, e.g., \cite{FH90}). So, the main increase in the blood
perfusion rate $j$ should fall on the temperature variations from $T_{a}$ to
a certain value $T_{\text{vr}}\approx T_{+}$ and after the temperature
exceeds the value $T_{\text{vr}}$ the blood perfusion rate $j$ is likely to
depend weakly on temperature because the blood vessels exhaust their ability
to expand.

In order to analyze the temperature evolution in living tissue the
dependence $j(T)$ obtained experimentally under practically uniform heating
of large tissue regions is typically used. However, whether such a
dependence holds when the temperature distribution becomes substantially
nonuniform is a question. Indeed, in this case due to the vessel extent the
dependence $j(T)$ can alter not only its particular form but also become
functional. In other words, when the tissue is heated substantially
nonuniform the blood perfusion rate $j(\mathbf{r})$ at a given point $%
\mathbf{r}$ is determined, in general, by the whole temperature distribution
$\{T( \mathbf{r}^{\prime })\}$ over a certain neighborhood of this point
(concerning a similar behavior of the tissue response to variations in CO$%
_{2}$ concentration see, e.g., \cite{Mch89}). Specific details of the
mechanism by which living tissue responds to local temperature variations on
scales about 1~cm is also a question \cite{Mch89}. However, under strong
heating localized on such scales thermal selfregulation can be effectively
implemented through the response of microcirculatory bed to reduced O$_{2}$
partial pressure or increased concentration of metabolism products (e.g. CO$%
_{2}$, H$^{+}$, ADP, \textsl{etc.}) because higher temperatures result in a
metabolism intensification with a higher O$_{2}$ consumption. In addition,
this explains a possible delay of the tissue response to temperature
variations and enables the corresponding delay time to be estimated at
several minutes (see, e.g., \cite{Mch89}).

In order to describe the local thermoregulation we need to specify how each
vessel responds to the corresponding piece of information characterizing the
state of the tissue, in particular, its temperature. So, in deriving the
equation relating the blood perfusion rate $j(\mathbf{r},t)$ to the tissue
temperature $T(\mathbf{r},t)$ we make use of the following physiological
data \cite{Mch89}. The local selfregulation of blood perfusion in living
tissue on scales about one centimeter is due to expansion or contraction of
blood vessels making up a single microcirculatory bed and blood
redistribution over this vascular network is mainly controlled by a large
group of arteries differing in length significantly. The reaction of the
microcirculatory bed is governed by receptors responding to variations in CO$%
_{2}$ partial pressure, H$^{+}$ concentration, or other metabolism products.
However, as mentioned above, under strong heating such receptors can play
the role of effective thermoreceptors supplying the microcirculatory bed
with the needed information. Obviously, none of the vessels receives the
whole information on the tissue state, so there must be a certain
cooperative mechanism of information self-processing by which the behavior
of different vessels is consistent with each other in the manner enabling
the tissue to respond properly.

We have shown \cite{BOOK,LGenv95,LGdan96,Siam} that such a cooperative
mechanism of self-regulation can be implemented through the vessel response
to the blood temperature in the corresponding veins. The receptors mentioned
above are located directly in the cellular tissue as well as embedded into
the walls of vessels, including veins \cite{Mch89}. Those embedded into the
vessel walls are able to measure the concentration of the metabolism
products directly in blood and, thus, to measure effectively its
temperature. For small vessels (arteries and veins) the position of the
receptors governing their behavior is not a factor. This allows us to make
use of the proposed cooperative mechanism of self-regulation in the
description of tissue response to local and strong heating.

Under the adopted assumptions it turns out that for the normal tissue the
dependence of blood perfusion rate $j$ on the tissue temperature $T$ can be
approximately described by a local equation relating the values of $j(%
\mathbf{r})$ and $T(\mathbf{r})$ taken at the same point $\mathbf{\ r}$
until the temperature comes close to the vital boundary $T_{+}$. This is due
to the cooperative mechanism of selfregulation which involves the individual
response of each vessel to the corresponding piece of information and the
coordination of their behavior by the selfprocessing of information. The
adequate selfprocessing of information is implemented through heat
conservation during blood motion inside relatively large veins of the
microcirculatory bed. The vascular network whose behavior is governed by
this mechanism can supply each region of the cellular tissue with blood at
such a rate that is required of its individual demand and different regions
of the cellular tissue do not interfere with one other. The obtained
equation for the tissue response is of the form \cite
{BOOK,LGenv95,LGdan96,Siam}:
\begin{equation}
\tau _{v}\frac{\partial j}{\partial t}+j\frac{T_{+}-T}{T_{+}-T_{a}}=j_{0}
\label{ThC:e1.14}
\end{equation}
where $\tau_{v}$ is the delay time of tissue response and $j_{0}$ is the
blood perfusion rate provided the tissue is not affected. This result
practically holds for living tissue containing also a small necrosis domain
\cite{BOOK,AGL94} because the temperature of blood in veins whose length
exceeds its size by several times is not sensitive to the presence of
necrosis domain.

In the next subsection we analyze the models of local thermal coagulation
used previously by other authors and discuss the basic theoretical problems
met in such approaches.

\subsection{Distributed description of thermal coagulation\label%
{ThC:subsec.DM}}

The effect of heat diffusion on local thermal coagulation has been
considered and numerically simulated by a number of authors (see, e.g. \cite
{Sv85,JM95,MR95} and references therein). Leaving aside the description of
laser light propagation in the tissue we can generalize their models for the
necrosis formation to the following coupled equations for the temperature
field $T(\mathbf{r},t)$ and the field $\zeta (\mathbf{r},t)$ determining the
fraction of undamaged tissue at a given point $\mathbf{r}$ and time $t$:
\begin{eqnarray}
c\rho \frac{\partial T}{\partial t} &=&\nabla \left( \kappa _{\text{eff}
}\nabla T\right) -fc_{b}\rho _{b}j(T-T_{a})+q\,,  \label{ThC:e1.1} \\
\frac{\partial \zeta }{\partial t} &=&-\zeta \omega (T)\,.  \label{ThC:e1.2}
\end{eqnarray}
The thermal coagulation rate $\omega (T)$ depends strongly on the tissue
temperature $T$ and for typical values of temperature attained during the
treatment course can be approximated by the expression
\begin{equation}
\omega (T)=\omega _{0}\exp \left[ \frac{T-T_{0}}{\Delta }\right] \,.
\label{ThC:e1.3}
\end{equation}
Here $\omega _{0}=\omega (T_{0})$, where $T_{0}$ is a certain fixed
temperature and $\Delta $ is a constant. Expression~(\ref{ThC:e1.3}) is
actually a convenient approximation of the Arrhenius dependence $\omega
(T)\propto \exp\{-\frac{E }{T}\}$ inside a not wide temperature interval
governing the tissue coagulation. The available experimental data \cite{J94}
for the temperature dependence of the exposure time enable us to estimate
the value of $\Delta $ as $\Delta \sim 3-5\,^{\circ}$C ($\Delta \simeq
3.6\,^{\circ}$C for pig liver at $T_{0}=65\,^{\circ}$C). Below in numerical
calculations the dependence~(\ref{ThC:e1.3}) will be taken in the form
\begin{equation}
\omega (T)=0.2\exp \left[ \frac{T-60}{3.6}\right] ~\text{min}^{-1}\,,
\label{ThC:e1.3a}
\end{equation}
where the temperature $T$ is in degrees Celsius.

In what follows the system of equations~(\ref{ThC:e1.1}), (\ref{ThC:e1.2}),
and expression~(\ref{ThC:e1.3}) will be referred to as the ``distributed
model'' for local thermal coagulation. This model in fact is rather a
phenomenological one and should be modified in order to allow for
sophisticated effects revealing themselves in local coagulation.

Let us consider the properties of local thermal coagulation that, in fact,
make the mathematical description of this process stubborn and give rise to
the inconsistency of the distributed model. During a typical course of
thermotherapy treatment high temperatures of order of $T_{\text{max }}\sim
100^{\circ}$C are attained at the necrosis center, whereas at distant points
$T=T_{a}\approx 37~^{\circ}$C. Under such conditions the damage rate $\omega
(T)$ or its reciprocal value $t_{\text{cg}}(T)=1/\omega (T)$ called the
threshold exposure time of thermal coagulation varies extremely sharp in
space on scales of the necrosis size $\Re $. Indeed, at points where the
temperature attains, for example, the values of $T=60~^{\circ}$C, $65~^{
\circ}$C, $70~^{\circ}$C, and $75~^{\circ}$C the values of $t_{\text{cg}}$
are $8$~min, $100$, $20$, and $5$~sec, respectively \cite{J94}. The typical
duration $t_{\text{course}}$ of the thermotherapy course is about several
minutes and estimated by the inequality, $t_{\text{course}}\gtrsim \tau $
\cite{we1,we3}, where
\begin{equation}
\tau \overset{\text{def}}{=}\frac{1}{f\left<j\right>}\sim 3-6~\text{min},
\label{ThC:e1.15}
\end{equation}
is the characteristic time scale of the necrosis growth and $\left<j\right>$
is the mean value of the blood perfusion rate attained near the necrosis
boundary. In estimate~(\ref{ThC:e1.15}) we have set $f=0.5$ and used the
typical values of the blood perfusion rate $\left<j\right>\sim 0.3$, $0.5$,
and $0.7$~min$^{-1}$ for stomach, intestine, and spleen, respectively~\cite
{BZK90}. So the layer wherein the thermal coagulation is under way at a
given moment of time is characterized by a narrow temperature interval of
thickness $\Delta $ located in the vicinity of the temperature of $T_{\text{%
cg}}\sim 65\,^{\circ}$C. This layer $\mathbb{L}_{\text{pd}}$ of partially
damaged tissue separates the necrosis region ($\zeta \ll 1$) and practically
live tissue ($\zeta \simeq 1$ ). Thus, in order to describe the necrosis
growth due to local thermal coagulation in the framework of the distributed
model one has to consider in detail the temperature field inside the layer $%
\mathbb{L}_{\text{pd}}$, because exactly it governs the necrosis growth
directly.

Now we estimate the thickness $\delta _{\text{pd}}$ of this layer. As
follows from the numerical analysis \cite{we1,we3} the temperature
distribution is characterized by a single spatial scale, so, in the necrosis
domain the mean temperature gradient is about $\left<\nabla T\right>\sim (T_{%
\text{max}}-T_{a})/\Re ${. In addition, the necrosis size attaining
typically the values of }$10-20${~mm meets the inequality~\cite{we1,we3}: }
\begin{equation}
\Re \gtrsim \ell _{T}\sim 10\,\text{mm}  \label{ThC:e1.16}
\end{equation}
by virtue of (\ref{ThC:e1.7}) and (\ref{ThC:e1.7a}){.} Then setting $\delta
_{\text{pd}}\left<\nabla T\right>\sim \Delta $ we get the desired estimate:
\begin{equation}
\delta _{\text{pd}}\sim \frac{\Delta \sqrt{L_{n}}}{(T_{\text{max}}-T_{a})}
\ell _{v}\sim 0.2\,\ell _{v}\sim 1~\text{mm\thinspace }.  \label{ThC:e1.17}
\end{equation}
So the layer $\mathbb{L}_{\text{pd}}$ of partially damaged tissue is
sufficiently thin, its thickness is less not only than the mean necrosis
size $\Re $ but also, what is essential for the theory development, than the
characteristic length $\ell _{v}$ of the vessels directly governing the heat
exchange between the cellular tissue and blood. Therefore, in order to
describe rigorously the temperature distribution on scales about $\delta _{%
\text{pd}}$ we have to go beyond the framework of the mean field
approximation. However, the distributed model is, in fact, a mean field one,
because it regards living tissue in terms of an effective continuum. Thereby
the distributed model in its own right cannot ensure the reliability for the
mathematical modeling of local tissue coagulation.

In particular, as an aspect of this problem we note that the mean field
approximation ignores the difference between the true temperature
distribution in the tissue and the temperature field $T(\mathbf{r},t)$, i.e.
does not take into account the random component of temperature distribution
caused by the vessel discreteness. This is justified if the mean amplitude $%
\sigma $ of such temperature nonuniformities is sufficiently small. By
virtue of (\ref{ThC:e1.7}) and setting the temperature inside the layer $%
\mathbb{L}_{\text{pd}}$ of partially damaged tissue equal to $T_{\text{cg}
}\approx 65\,^{\circ}$C we get the estimate
\begin{equation}
\left. \sigma \right| _{\mathbb{L}_{\text{pd}}}\sim \Delta \,.
\label{ThC:e1.18}
\end{equation}
In other words, inside the layer $\mathbb{L}_{\text{pd}}$ the amplitude of
the temperature nonuniformities due to the vessel discreteness turns out to
be about the characteristic temperature drop across the whole layer $\mathbb{%
L}_{\text{pd}}$. So such temperature nonuniformities are bound to affect the
necrosis formation.

Another problem inherent in the distributed model is the specification of
the main kinetic coefficients characterizing the partially damaged tissue as
a continuous medium that effectively approximates the real cellular tissue
with the embedded vascular network, with both of them having undergone
partial damage. Namely, we should specify the effective thermal conductivity
$\kappa _{\text{eff}}$ and the factor $f$ as functions of the blood
perfusion rate $j$ and the fraction $\zeta $ of undamaged tissue:
\begin{equation}
\kappa _{\text{eff}}=\kappa _{\text{eff}}(\zeta ,j)\,,\qquad f=f(\zeta ,j)\,.
\label{ThC:e1.19}
\end{equation}
However, these coefficients being the characteristics of the mean field
theory can adequately describe heat transfer in living tissue only on scales
about or greater than $\ell _{v}$. So we may write the corresponding
relationships only for the live tissue where $\zeta \approx 1$. (Clearly,
inside the necrosis domain $\kappa _{\text{eff}}=\kappa $ and we may set $f$
equal to any value of order unity because of $j=0$ in this region.) Inside
the layer of partially damaged tissue any dependence of the type $\kappa _{%
\text{eff}}=\kappa _{\text{eff}}(\zeta )$ or $f=f(\zeta )$ is nothing more
than a formal phenomenological approximation of a real complicated
phenomenon. In particular, as follows from the results to be obtained the
particular details of the dependence $f=f(\zeta )$ are of no concern and
below we will treat the value $f$ as a constant.

Besides, in order to complete the distributed model one should describe how
the blood perfusion rate depends on the undamaged tissue fraction $\zeta $
and the temperature. It has been proposed \cite{JM95} to set
\begin{equation}
j(\zeta ,T)=\zeta j_{t}(T)\,,  \label{ThC:e1.20}
\end{equation}
where $j_{t}(T)$ is the perfusion rate that would occur in tissue without
damage. Clearly, it is also a pure phenomenological approximation.

The only fact, that is able to justify the feasibility of applying the
distributed model to the analysis of local thermal coagulation can be the
independence of the necrosis growth from the particular details of heat
transfer in the layer of partially damaged tissue. In this case, however, it
would be more consistent to use a free boundary model which ignores the
thickness of this layer, i.e. treats it as the boundary of the necrosis
domain. The motion of such an interface must be governed by the boundary
values of the temperature and its gradient. Particular details of heat
transfer in the given layer may be taken into consideration by a certain
collection of parameters.

In the next section we will show that this independence is the case and
derive the corresponding free boundary model. In particular, formally
assuming equation~(\ref{ThC:e1.1}) to hold in the whole region under
consideration we will reduce the system of equations~(\ref{ThC:e1.1}), (\ref
{ThC:e1.2}) to a certain equivalent free boundary model dealing with
integrated characteristics of the tissue damage rate inside the layer $%
\mathbb{L}_{\text{pd}}$. In other words, the model to be obtained aggregates
the particular details of the functions $\kappa _{\text{eff} }(\zeta )$ and $%
j(\zeta ,T)$ to certain contains of order unity, which demonstrates the
desired independence.

\section{Interface dynamics\label{ThC:sec.ID}}

\begin{figure}[h]
\begin{center}
\includegraphics[width=60mm]{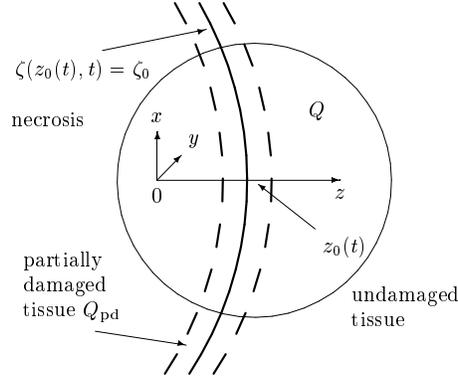}
\end{center}
\caption{The layer $\mathbb{L}_{\text{ pd}}$ of partially damaged tissue and
the local coordinate system.}
\label{ThC:F2}
\end{figure}

In order to obtain the free boundary description of local thermal
coagulation it is suffice to consider a certain neighborhood $Q$ of the
partially damaged tissue layer $\mathbb{L}_{\text{pd}}$ (Fig.~\ref{ThC:F2}).
The size of the region $Q$ is assumed to be, on one hand, much smaller than
the characteristic size $\Re $ of the necrosis domain and, on the other
hand, much larger than the thickness $\delta _{\text{pd}}$ of the layer $%
\mathbb{L}_{\text{pd}}$. In other words, we analyze the case when the tissue
in the region affected directly by laser light has already coagulated and
the necrosis growth is caused by the heat diffusion into the surrounding
tissue. To describe the dynamic of local thermal coagulation we introduce
the interface $\Gamma$ specified by the condition
\begin{equation}
\left. \zeta (\mathbf{r},t)\right| _{\mathbf{r}\in \Gamma }=\zeta _{0},
\label{ThC:e2.0}
\end{equation}
where $\zeta _{0}\sim 0.5$ is a certain fixed value, and keep track of how
it moves. Then the necrosis growth is represented by the motion of the
interface $\Gamma $ at the normal velocity $\vartheta_n$.

There are two ways of deriving the desired expression for the velocity $%
\vartheta _{n}$. One of them is as follows. Let us, first, assume equations~(%
\ref{ThC:e1.1}), (\ref{ThC:e1.2}) to hold at all the points of the tissue.
Then the derivation is to reduce the distributed model to certain equations
dealing with the layer $\mathbb{L}_{\text{pd}}$ of partially damaged tissue
in terms of an interface $\Gamma$ whose velocity $\vartheta_n$ is determined
by the local characteristics of the temperature field that practically do
not vary over this layer and so remains unchanged at the nearest points of
the live tissue and the necrosis region. This way has been implemented in
our papers \cite{we2,we6} using the singular perturbation technique in the
small parameter $\Delta /(T_{\text{max}}-T_{a}) $. In particular, we have
obtained the following expression relating the velocity $\vartheta _{n}$ of
the interface $\Gamma$, the value $T_{\text{cg} }$ of the temperature at
this interface, and the boundary value of the temperature gradient, for
example, on the necrosis side $\left. \nabla _{n}T\right| _{\Gamma -0}$:
\begin{equation}
\vartheta _{n}=\Im _{0}\frac{\Delta }{\left| \nabla _{n}T\right| _{\Gamma
-0} }\omega (T_{\text{cg}})\,,  \label{ThC:e2.1}
\end{equation}
where the function $\omega (T)$ is the thermal coagulation rate given by
formula~(\ref{ThC:e1.3}) and the constant $\Im _{0}$ takes into account the
formal dependence of the effective thermal conductivity $\kappa _{\text{eff}
}(\zeta )$ on the undamaged tissue fraction $\zeta $
\begin{equation}
\frac{1}{\Im _{0}}=\int\limits_{\zeta _{0}}^{1}d\zeta \,\frac{\kappa }{
\kappa _{\text{eff}}(\zeta )\zeta }\sim 1\,.  \label{ThC:e2.2}
\end{equation}

The other way of obtaining expression~(\ref{ThC:e2.1}) is to state it basing
on the general properties of thermal coagulation and without using the
distributed model at all. Doing so we are not be able to get an expression
for the constant $\Im _{0}$ and, thereby, should regard it as a
phenomenological one. However, first, expression~(\ref{ThC:e2.2}) is as
rigorous as the former approach using the dependence $\kappa _{\text{eff}%
}(\zeta )$. Second, following the latter way we show that the free boundary
model based on expression~(\ref{ThC:e2.1}) is more general than the
distributed model, i.e. it is not restricted to the validity area of the
distributed model.

Let thermal coagulation be under way in a layer $\mathbb{L}_{\text{pd}}(t)$
at a given moment of time $t$ and the mean temperature and the mean
temperature gradient inside this layer be $T_{\text{cg}}$ and $\nabla_{n}T_{%
\text{pd}}$, respectively. Then the characteristic thickness $\delta_{\text{%
pd}}$ of the layer $\mathbb{L}_{\text{pd}}$ can be estimated by expression
\begin{equation}
\delta _{\text{pd}}\sim \frac{\Delta }{\nabla _{n}T_{\text{pd}}}\,.
\label{ThC:e2.3}
\end{equation}
In fact, when the necrosis growth due to thermal coagulation is limited by
the heat diffusion the temperature distribution is substantially nonuniform
in space. Therefore, on one hand, in the necrosis region, where the
temperature exceeds $T_{\text{cg}}$ by values greater than $\Delta $: ($T-T_{%
\text{cg} }\gtrsim \Delta $) and, thus, $\omega (T)\gg \omega (T_{\text{cg}%
}) $, the tissue coagulation has to be completed. On the other hand, in the
region of undamaged tissue the temperature is sufficiently low, $T_{\text{cg}
}-T\gtrsim \Delta $, and, so, $\omega (T)\ll \omega (T_{\text{cg}})$,
thereby the tissue has not enough time to coagulate at such temperatures.
Therefore tissue coagulation can be under way only in the region where $%
\left| T-T_{\text{cg}}\right| \lesssim \Delta $ whence we immediately get
estimate~(\ref{ThC:e2.3}). After a lapse of the time interval $t_{\text{ thr}
}\sim 1/\omega (T_{\text{cg}})$ the tissue in the layer $\mathbb{L}_{\text{pd%
}}$ has to coagulate practically completely. This is equivalent to the
displacement of the layer $\mathbb{L}_{\text{pd}}$ over the distance $%
\delta_{\text{pd}}$. Thereby, if we keep watch on the points at which $\zeta
\sim 0.5 $ then we will see that these points move at the velocity
\begin{equation*}
\vartheta \sim \frac{\delta _{\text{pd}}}{t_{\text{ thr}}}\sim \frac{\Delta
}{\nabla _{\text{pd}}T}\omega (T_{\text{cg}})\,,
\end{equation*}
which exactly coincides with expression~(\ref{ThC:e2.1}) within a factor of
order unity.

The other relation completing the free boundary approximation of the
necrosis growth reflects the heat flux conservation through the layer $%
\mathbb{L} _{\text{pd}}$ of partially damaged tissue. Obviously, it can be
represented in the form
\begin{equation}
\kappa \left. \nabla _{n}T\right| _{\Gamma -0}=\kappa _{\text{eff}}\left.
\nabla _{n}T\right| _{\Gamma +0}\,  \label{ThC:e2.4}
\end{equation}
relating the temperature gradients $\left. \nabla _{n}T\right| _{\Gamma -0}$
and $\left. \nabla _{n}T\right| _{\Gamma +0}$ at the interface $\Gamma $ on
the necrosis side and the side of live tissue, respectively.

Expressions~(\ref{ThC:e2.1}), (\ref{ThC:e2.4}) are actually the essence of
the free boundary model for the dynamics of local thermal coagulation. It
should be pointed out that the stated free boundary approximation in its
turn can be regarded as the initial point of modeling the necrosis growth
due to local thermal coagulation. Indeed, all the information required of
specifying the dependence $\vartheta _{n}(T)$ can be obtained based on the
experimental data for the temperature dependence of the threshold exposure
time at fixed temperature. The only parameter of this model that contains
the particular information about the properties of heat transfer in the real
layer of partially damaged tissue is the numeric factor $\Im _{0}$ of order
unity, $\Im _{0}\sim 1$. Keeping the latter in mind we may regard the
analysis presented in the given section as the substantiation of the fact
that the necrosis growth due to local thermal coagulation limited by heat
diffusion is insensitive to the particular details of heat transfer inside
partially damaged tissue. Thus, it actually justifies the distributed model
rather than the free boundary model.

Now the obtained results (\ref{ThC:e2.1}), (\ref{ThC:e2.4}) enable us to
state the desired free boundary model for heat diffusion limited thermal
coagulation.

\section{Free boundary model\label{11}}

\begin{figure}[tbp]
\par
\begin{center}
\includegraphics[width=60mm]{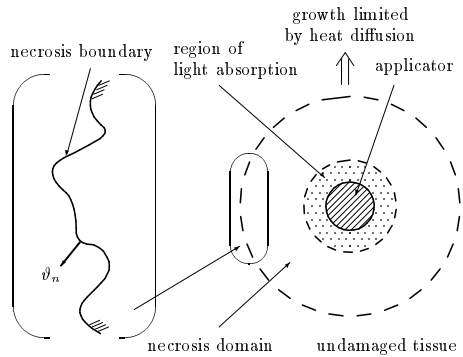}
\end{center}
\caption{The necrosis growth due to local thermal coagulation limited by
heat diffusion. The physical system under consideration.}
\label{ThC:F3}
\end{figure}

Following the previous section the tissue is considered as comprising two
regions (Fig.~\ref{ThC:F3}), the necrosis domain $Q_{\text{n}}$ where the
blood perfusion rate is equal to zero,
\begin{equation}
j(\mathbf{r},t)=0\quad \text{for}\quad \mathbf{r}\in Q_{\text{n}}\,,
\label{ThC:e3.1}
\end{equation}
and the undamaged tissue $Q_{\text{t}}$ responding to temperature variations
by increasing locally the blood perfusion rate $j(\mathbf{r},t)$. Besides,
we allow for that the tissue response can be delayed. Blood vessels can
expand only to a certain extent as the temperature grows. So when it becomes
high enough, $T>T_{\text{vr}}$, the blood perfusion rate $j(\mathbf{r},t)$
attains a large but finite value $j_{\text{max}}$ and remains approximately
constant. Taking into account expression~(\ref{ThC:e1.14}) we describe this
behavior of the undamaged tissue by the equation:
\begin{equation}
\tau _{v}\frac{\partial j}{\partial t}+j\Phi (T)=j_{0}\quad \text{for}\quad
\mathbf{r}\in Q_{\text{t}}\,.  \label{ThC:e3.2}
\end{equation}
Here $\tau _{v}$ is the delay time of the tissue response and the function $%
\Phi (T)$ is of the form
\begin{equation}
\Phi (T)=\left\{
\begin{array}{ll}
\alpha +(1-\alpha ){\frac{T_{\text{vr}}-T}{T_{\text{vr}}-T_{a}}} & \quad
\text{for}\quad T<T_{\text{vr}} \\
\alpha & \quad \text{for}\quad T>T_{\text{vr}}
\end{array}
\right.  \label{ThC:e3.3}
\end{equation}
where $\alpha =j_{0}/j_{\text{max}}$ is the ratio of $j_{0}$ and the maximum
$j_{\text{max}}$ of the blood perfusion rate that can be attained in living
tissue due to the vessel expansion caused by the tissue response to
temperature increase, and $T_{\text{vr}}\approx 45$--$46^{\circ}$C is the
temperature at which the blood vessels exhaust their ability to expand.
Inside the necrosis domain $Q_{\text{n}}$ the tissue temperature obeys the
heat diffusion equation for solids:
\begin{equation}
c\rho \frac{\partial T}{\partial t}=\kappa \nabla ^{2}T+q  \label{ThC:e3.4}
\end{equation}
where $\kappa $ is the cellular tissue conductivity and $q$ is the rate of
heat generation due to the laser light absorption. Inside the undamaged
tissue the temperature is governed by the equation taking into account also
the effect of the vessel discreteness (replacement~(\ref{ThC:e1.12})):
\begin{equation}
c\rho \frac{\partial T}{\partial t}=F\kappa \nabla ^{2}T-fc_{b}\rho
_{b}(j_{v}+\delta j)(T-T_{a})+q\,.  \label{ThC:e3.5}
\end{equation}
Here, as before, $j_{v}$ is the regular component of the blood perfusion
rate averaged over spatial scales of order $\ell _{v}$, the value $\delta j$
is its component due to the vessel discreteness treated as random, and the
constants $F=\kappa _{\text{eff}}/\kappa >1$ and $f<1$ are of order unity.
At the interface $\Gamma $ between the necrosis domain and the undamaged
tissue the temperature and the heat flux are assumed to have no jumps, i.e.
the temperature distribution meets the following boundary conditions
\begin{equation}
\left. T\right| _{\Gamma +0}=\left. T\right| _{\Gamma -0}\overset{\text{def}
}{=}T_{\text{cg}}\,,\qquad \left. F\nabla _{n}T\right| _{\Gamma +0}=\left.
\nabla _{n}T\right| _{\Gamma -0}\,.  \label{ThC:e3.6}
\end{equation}
By virtue of (\ref{ThC:e1.3}) and (\ref{ThC:e2.1}) the normal velocity of
the interface $\Gamma $ is given by the expression
\begin{equation}
\vartheta _{n}=\frac{\Im _{0}\omega _{0}\Delta }{\left| \nabla _{n}T\right|
_{\Gamma -0}}\exp \left[ \frac{T_{\text{cg}}-T_{0}}{\Delta }\right] \,
\label{ThC:e3.7}
\end{equation}
For points distant from the necrosis domain
\begin{equation}
\left. T\right| _{\infty }=T_{a}.  \label{ThC:e3.8}
\end{equation}

Finding the relationship between the averaged and true blood perfusion
rates, $j_{v}$ and $j$, we have to take into account that the scale $%
\ell_{v} $ of averaging in its turn depends on the local value of $j_{v}$ as
demonstrated by expression~(\ref{ThC:e1.4}). This dependence enables us to
specify functional~(\ref{ThC:e1.6}) in the form \cite{BOOK}:
\begin{equation}
j_{v}-\frac{\lambda_v \kappa }{c\rho }\nabla ^{2}\ln j_{v}=j\,,
\label{ThC:e3.9}
\end{equation}
where $\lambda_v \sim 1/\sqrt{L_{n}}$ is also a constant of order unity.
Equation~(\ref{ThC:e3.9}) should be subjected to a certain boundary
condition at the interface $\Gamma $ because it makes no sense to average
the blood perfusion rate over the necrosis domain impermeable to blood. The
physical layer separating the necrosis domain and the undamaged tissue where
the local vascular network is not damaged is complex in structure and
contains a spatial increase of the blood perfusion rate from zero to the
value in the undamaged tissue. In order to avoid the problem of analyzing
the blood perfusion rate in this layer we take into account the following
simplifying circumstance. On one hand, the typical size of the necrosis
domain formed during a thermal therapy course and the characteristic length
of temperature variations are of the same order about 1~cm. So, particular
details of spatial variations in the blood perfusion rate on scale much less
than 1~cm are not the factor. On the other hand, the damaged part of the
vascular network located inside the necrosis domain is most probable to be
made up of an artery and vein having supplied previously this region with
blood as a whole and of shorter vessels formed by their branching. Indeed,
the mean volume of living tissue falling on a single small artery (or vein)
of a fixed length $l$ is about $l^{3}$ \cite{Mch89,AAS92}. Thus, the regions
where the total blood perfusion is directly controlled by different vessel
of fixed length do not cross each other substantially and the architectonics
of microcirculatory bed can be approximately represented as the binary tree
embedded uniformly into the cellular tissue \cite{AAS92}. Therefore, the
region containing the vascular network part in which blood flow is
remarkably disturbed because of the necrosis formation does not exceed
substantially the necrosis domain. The latter enables us not to make
difference between the given layer and the interface $\Gamma $ and to deal
with a sharp jump of the blood perfusion rate at the necrosis interface. The
desired boundary condition imposed on the averaged blood perfusion rate $%
j_{v}$ must obey the law of blood conservation, which in this case can be
written as
\begin{equation}
\int\limits_{Q_{\text{t}}}d\mathbf{r\,}j_{v}(\mathbf{r},t)=\int\limits_{Q_{%
\text{t}}}d\mathbf{r\,}j(\mathbf{r},t)\,.  \label{ThC:e3.10}
\end{equation}
So in order to fulfill identity (\ref{ThC:e3.9}) we have to set the normal
gradient of the averaged blood perfusion rate equal to zero at the interface
$\Gamma $:
\begin{equation}
\left. \nabla _{n}j_{v}\right| _{\Gamma ^{+}}=0\,.  \label{ThC:e3.11}
\end{equation}
We note that the adopted assumption will not hold if a large vessel goes
through the necrosis domain. However, the probability of this event is small
enough and this case should be analyzed individually.

\begin{figure}[h]
\begin{center}
\includegraphics[width=55mm]{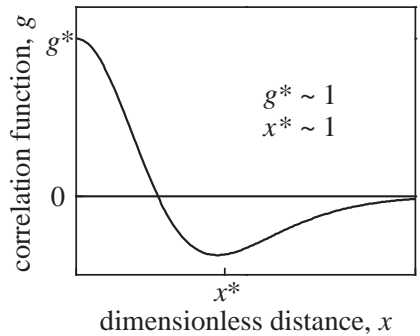}
\end{center}
\caption{Typical form of the correlation function of the blood perfusion
rate nonuniformities caused by the vessel discreteness.}
\label{ThC:F4}
\end{figure}

The random component $\delta j$ of the perfusion rate allowing for the
vessel discreteness on scales $\ell_{v}$ is assumed to meet the conditions:
\begin{eqnarray}
\left\langle \delta j(\mathbf{r},t)\right\rangle &=&0\,,  \label{ThC:e3.12a}
\\
\left\langle \delta j(\mathbf{r},t)\delta j(\mathbf{r}^{\prime
},t)\right\rangle &\simeq &j_{v}(\mathbf{r},t)j_{v}(\mathbf{r}^{\prime
},t)g( \mathbf{r},\mathbf{r}^{\prime })\,.  \label{ThC:e3.12b}
\end{eqnarray}
Here the correlation function $g(\mathbf{r},\mathbf{r}^{\prime })$ is
symmetric with respect to the exchange of the variables $r$, $r^{\prime }$,
i.e. $g(\mathbf{r},\mathbf{r}^{\prime })=g(\mathbf{r}^{\prime },\mathbf{r})$
and has to obey the general law of blood conservation:
\begin{multline}
\left< \delta j(\mathbf{r},t) \int\limits_{Q_{\text{t}}}d\mathbf{r}%
^{\prime}\,\delta j(\mathbf{r}^{\prime},t) \right> = {} \\
j_v(\mathbf{r},t)\int\limits_{Q_{\text{t}}}d\mathbf{r}^{\prime}\,j_v(\mathbf{%
r}^{\prime},t) g(\mathbf{r},\mathbf{r}^{\prime})=0\,.  \label{ThC:e3.14}
\end{multline}
For the living tissue without necrosis it is natural to assume:
\begin{equation}
g(\mathbf{r},\mathbf{r}^{\prime })=g^{0}\left( \frac{\left| \mathbf{r-r}
^{\prime }\right| }{\ell _{v}}\right) \,,  \label{ThC:e3.15}
\end{equation}
where the function $g^{0}(x)$ is such that $g(x)\sim 1$ for $x\lesssim 1$, $%
\left|g(x)\right|\ll 1$ for $x\gg 1$, and
\begin{equation}
\int\limits_{0}^{\infty }dx\,x^{2}g^{0}(x)=0\,.  \label{ThC:e3.160}
\end{equation}
The latter reflects the blood conservation in three-dimensional space. A
typical form of the function $g^{0}(x)$ is shown in Fig.~\ref{ThC:F4}

The system of equations (\ref{ThC:e3.1}), (\ref{ThC:e3.2}), (\ref{ThC:e3.4}%
), (\ref{ThC:e3.5}), (\ref{ThC:e3.9}) subject to the boundary conditions~(%
\ref{ThC:e3.6})--(\ref{ThC:e3.8}), (\ref{ThC:e3.11}) with expressions~(\ref
{ThC:e3.12a}), (\ref{ThC:e3.12b}) makes up the desired free boundary model.

Keeping in mind that the distributed model was widely used let us, now,
compare the dynamics of the necrosis growth predicted with this model and
with the stated free boundary model under different physical conditions. In
this way, first, we will verify the basic question as to whether the value $%
\Delta $ of the order of $3-5~^{\circ}$C is small enough so the ratio $%
\Delta /(T_{\text{max}}-T_{a})$ may be regarded as a sufficiently small
parameter for the free boundary model to hold. Second, we will actually
justify the distributed model because, as will be seen, it does be
independent of the particular details of heat transfer inside the layer of
partially damaged tissue.

\subsection{Model comparison\label{ThC:subsec.MC}}

\begin{figure}[tbp]
\par
\begin{center}
\includegraphics[width=60mm]{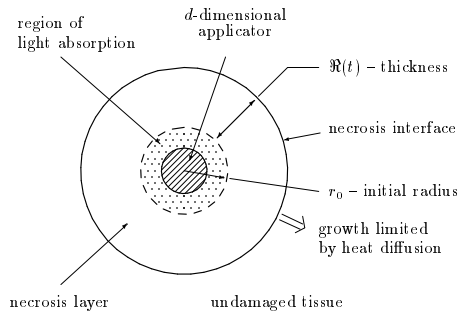}
\end{center}
\caption{The form of the necrosis region under consideration, $d={}$3, 2, 1
corresponds to applicator of the spherical, cylindrical, and plane form,
respectively (in the latter case $r_{0}\rightarrow \infty $).}
\label{ThC:F5}
\end{figure}

To be defined we simulate the necrosis growth in the tissue phantom shown in
Fig.~\ref{ThC:F5}. The applicator presented by the hatched circle locally
heats the tissue up to high temperatures of the order $T_{b}\sim 100$~$^{
\circ}$C, which causes immediate tissue coagulation in the nearest
neighborhood of the applicator. This can be the case, for example, because
of the direct heat exchange between the tissue and the applicator heated to
such temperatures or due to irradiation of laser light and its absorption in
an adjacent thin layer (dotted region in Fig.~\ref{ThC:F5}). In the latter
case it is typically used additional internal cooling of the applicator
boundary, so the maximum $T_{b}$ of the temperature attained just near the
applicator does not practically depend on heat diffusion into the
surrounding tissue. So keeping in mind the possible vaporization control
over the temperature maximum we will treat the value $T_{b}$ as a boundary
temperature fixed at the interface of a certain radius $r_{0}$. Generalizing
both these situations let us confine ourselves to the analysis of local
coagulation assuming that:

\begin{itemize}
\item[-]  at the initial time $t=0$ the necrosis region under consideration
is layer of zero thickness whose finite radius is $r_{0}$: $\zeta (r)=0$ and
$T(r)=T_{a}\approx 37$~$^{\circ }C$ for $r>r_{0}$;

\item[-]  the following necrosis growth is governed solely by heat
diffusion, i.e. we set $q(\mathbf{r},t)=0$\ for $r>r_{0}$;

\item[-]  at the boundary $r=r_{0}$ the temperature is a fixed value: $%
T(r_{0})=T_{b}\approx 100$~$^{\circ }C$;

\item[-]  at distant points the tissue temperature is equal to $T_{a}$: $%
T\rightarrow T_{a}$ as $r\rightarrow \infty $.
\end{itemize}

In other words, we confine ourselves to the region $r>r_0$ where the layer $%
r_0<r<r_0+\Re (t)$ of thickness $\Re (t)$ represents the necrosis domain
whose growth is directly governed by the heat diffusion only. The processes
in the region $r<r_0$ remain beyond consideration, i.e. we ignore the real
dynamics of initial coagulation in the immediate vicinity of the applicator
boundary. The typical duration of the latter process can be estimated as $%
1/\omega (T_b)\ll 1$~sec for the applicator directly heating the surrounding
tissue and as $\max [1/\omega (T_b),q(r_0)/(c\rho (T_b-T_a))]$ for the laser
applicator. In the present analysis this duration is regarded as a small
parameter.

Keeping in mind applicators of various forms we study the necrosis growth in
the one-, two-, and three-dimensional tissue phantoms. The tissue response
to temperature variations is described by equation~(\ref{ThC:e3.2}) and
expression~(\ref{ThC:e3.3}) with $\alpha =0.1$ because the tissue response
is sufficiently strong and the blood perfusion rate can locally increase by
tenfold \cite{Song84}. For the main tissue parameters we use the values
pointed out in Sec.~\ref{ThC:subsec.HM} which give us the following
estimates
\begin{equation*}
\tau _{0}=\frac{1}{fj_{0}}\approx 6\,\text{min}\quad \text{and}\quad \ell
_{0}=\sqrt{\frac{\kappa }{c\rho fj_{0}}}\approx 10\,\text{mm}
\end{equation*}
of the characteristic temporal and spatial scales, $\tau_0$ and $\ell_0$. We
compare the dynamics of the necrosis growth, namely, the time dependence of
the thickness $\Re (t)$ of the necrosis layer and the coagulation
temperature $T_{\text{cg}}(t)$ predicted by the developed free boundary
model with that given by the distributed model. In the latter case we use
expression~(\ref{ThC:e1.20}) and specified the required dependence of the
effective thermal conductivity $\kappa _{\text{eff}}$ on the fraction $\zeta
$ of undamaged tissue by the function
\begin{equation*}
\kappa _{\text{eff}}(\zeta )/\kappa =(F-1)\zeta +1\,.
\end{equation*}
Besides, for the distributed mode the necrosis interface $\Gamma$ is
specified by the condition $\left. \zeta \right|_{r\in \Gamma}=0.5$. For the
sake of simplicity in the model comparison we also ignore the effect of
vessel discreteness and identify the true and averaged blood perfusion rates.

\begin{figure}[tbp]
\par
\begin{center}
\includegraphics[width=60mm]{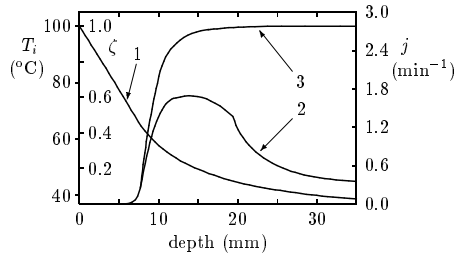}
\end{center}
\caption{The typical form of the spatial distribution of the tissue
temperature (curve 1), blood perfusion rate (curve 2), and fraction of
undamaged tissue (curve 3). (In obtaining the present curves we have set $j_{%
\text{max}}=10\,j_0$ and $\protect\tau_v \approx 1$~min and considered the
one-dimensional tissue phantom.)}
\label{ThC:F6}
\end{figure}

First of all, Fig. \ref{ThC:F6} demonstrates the typical form of the
temperature distribution $T(r)$, the distribution of the undamaged tissue
fraction $\zeta (r)$, and the blood perfusion rate $j(r)$ obtained for the
one-dimensional tissue phantom in the frames of the distributed model. As
seen in Fig. \ref{ThC:F6} the region wherein the undamaged tissue fraction $%
\zeta$ varies substantially in space is sufficiently thin. So on spatial
scales characterizing the temperature decrease such an increase of the value
$\zeta(r)$ may be treated as a sharp jump. The latter actually justifies
using the free boundary model for the given values of the tissue parameters,
namely, the capability for assigning to an effective necrosis boundary a
certain coagulation temperature $T_{\text{cg}}$ and certain values (on both
the sides) of the temperature gradient.

\begin{figure}[tbp]
\par
\begin{center}
\includegraphics[width=85mm]{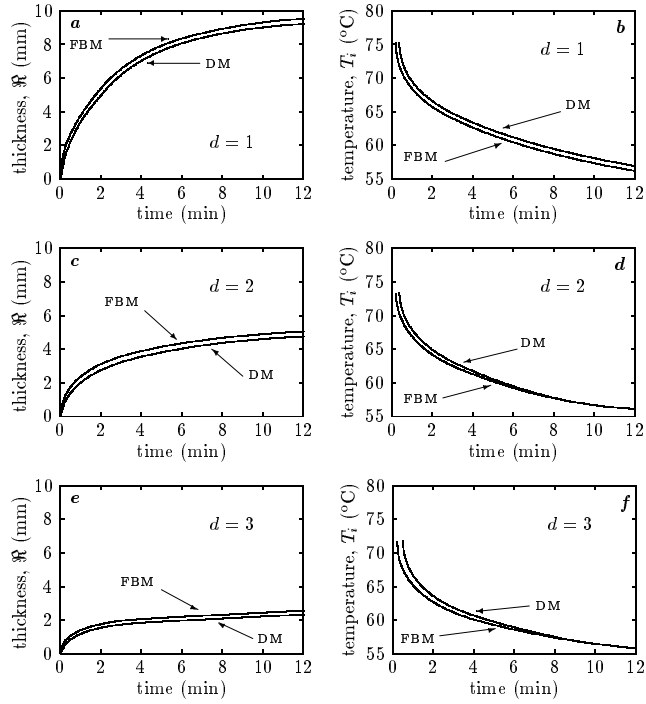}
\end{center}
\caption{Comparison of the necrosis growth predicted by the distributed
model (DM) and the free boundary model (FBM). The thickness $\Re(t)$ of the
necrosis layer and the temperature $T_i(t)$ at the necrosis interface vs
time $t$ for heat sources of the plane ($a$, $b$), cylindrical ($c$, $d$),
and spherical ($e$, $f$) form. (In numerical calculations we have set $%
\protect\tau_v = 2$~min, $j_{\text{max}} = 10 j_0$. For the distributed
model the value $T_i$ is specified as $T(r_i)$ at the point $r_i$ at which $%
\protect\zeta(r_i)=0.5$.)}
\label{ThC:F7}
\end{figure}

Fig.~\ref{ThC:F7} demonstrates the fact that the distributed model leads
practically to the same dynamics of the necrosis growth as that predicted by
the free boundary model for one-, two- , and three-dimensional tissue
phantom (figures ($a$, $b$), ($c$, $d$), and ($e$, $f$), respectively).
These results have been obtained for the tissue phantom with the strong ($j_{%
\text{max}}=10j_{0}$) and delayed ($\tau _v=2$~min) response to temperature
variations and so actually comprises the characteristic features of the
necrosis growth in the tissue without response as well as with a strong
immediate response.

The present figures not only clearly justify that the particular details of
thermal coagulation in a real partially damaged layer are not the factor but
also show that the ratio $\Delta /(T_{\text{max}}-T_{a})$ for $\Delta
\approx 3-5~^{\circ}$C can be treated in fact as a small parameter. So the
two models may be regarded formally as equivalent, but the free boundary one
does not contain self-inconsistent elements. In addition the free boundary
model can be applied to constructing a faster numerical algorithm of
simulating the necrosis growth because in this model we need not consider
the thin layer of partially damaged tissue. In the free boundary model we
deal with the temperature field only in the necrosis region and the region
of undamaged tissue where it is sufficiently smooth.

\begin{figure}[tbp]
\par
\begin{center}
\includegraphics[width=60mm]{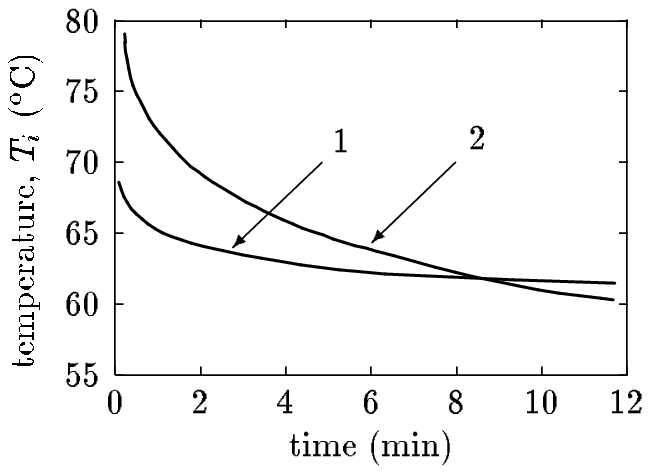}
\end{center}
\caption{The time dependence of the temperature $T_i$ at the point $r_i$
where $\protect\zeta(r_i) = 0.5$ for different values of the parameter $%
\Delta$. (Curves 1, 2 correspond to $\Delta = 1.5~{}^{\circ}$C and $\Delta =
5.0~{}^{\circ}$C, respectively. In obtaining the present curves we have
considered the one-dimensional phantom of the tissue without response to
temperature variation).}
\label{ThC:F8}
\end{figure}

In this analysis we have gotten another characteristic feature of local
thermal coagulation illustrated in Fig.~\ref{ThC:F8}. The free boundary
approximation is rigorously justified provided the temperature distribution
inside the layer of partially damaged tissue can be regarded as
quasistationary. The latter is the case when, in particular, the time
variations $\delta_{t}T_{\text{cg}}$ of the coagulation temperature $T_{%
\text{cg}}$ are small enough during the necrosis growth, which in
mathematical terms may be stated as the condition $\delta _{t}T_{\text{cg}
}\rightarrow 0$ as $\Delta \rightarrow 0$. Fig. \ref{ThC:F8} demonstrates
that this condition may be fulfilled. Indeed, the smaller is the parameter $%
\Delta $ of the tissue damaged rate $\omega (T,\Delta )$, to a greater
extent are smothered the time variations of the coagulation temperature $T_{%
\text{cg}}$ except for a short initial period of the necrosis growth.
Besides, the given feature of the local thermal coagulation justifies, at
least at the qualitative level, the a certain simplification of the free
boundary model \cite{we1,we3} that considers the coagulation temperature $T_{%
\text{cg}}$ fixed during the necrosis growth.

\subsection{Fixed coagulation temperature approximation\label{ThC:subsec.FT}}

In this approximation expression~(\ref{ThC:e3.7}) is replaced by the
conditions on the temperature at the necrosis boundary $\Gamma$. Namely, it
is assumed that inside the undamaged tissue the temperature cannot exceed
the coagulation temperature $T_{\text{cg}}$, i.e.
\begin{equation}
T(\mathbf{r},t)<T_{\text{cg}}\quad \text{for}\quad \mathbf{r}\in Q_{\text{t}}
\label{ThC:e3.17}
\end{equation}
and at the interface $\Gamma $ the boundary value $T_{i}$ is either equal to
the coagulation temperature, $T_{i}=T_{\text{cg}}$, or rigorously less: $%
T_{i}<T_{\text{cg}}$. The former case takes place when growing near the
interface $\Gamma $ the temperature comes closely to the value $T_{\text{cg}
} $ and the interface has to move in order to keep up the boundary
temperature $T_{i}$ inside the interval $T_{i}\leq T_{\text{cg}}$. In the
second case the interface is fixed. Both these conditions can be formally
described by the expression:
\begin{equation}
(T_{i}-T_{\text{cg}})(\left. \frac{\partial T}{\partial t}\right| _{\Gamma
}- \frac{\partial T_{i}}{\partial t})=0\,,  \label{ThC:e3.18}
\end{equation}
where the boundary temperature $T_{i}(\mathbf{s},t)$ is treated as a
function of the interface coordinates $\mathbf{s}$ and the time $t$.

This approximation, except for simplifying the mathematical analysis of
local coagulation, demonstrates the fact that the necrosis growth limited by
heat diffusion weakly depends on the particular details of thermal
coagulation. All the characteristics of thermal coagulation are aggregated
into the coagulation temperature $T_{\text{cg}}$ and the coagulation rate $%
\omega (T)$ is only required to depend on temperature sufficiently strong.

This fact is illustrated in the following way. We study the necrosis growth
in the one-dimensional tissue phantom within the framework of the
distributed model modified so to take into account the effect of the blood
perfusion nonuniformities near the necrosis domain. In other words, in
equation~(\ref{ThC:e1.1}) we replace the true perfusion rate by the averaged
one, $j\rightarrow j_{v}$, related by equation~(\ref{ThC:e3.9}) subject to
the boundary condition~(\ref{ThC:e3.11}). As before, the effect of vessel
discreteness is ignored. We keep track of the points $x_{0.2}$, $x_{0.5}$, $%
x_{0.8}$ specified by the equalities $\zeta =0.2$,$~0.5$,$~0.8$,
respectively. In this way the dynamics of the necrosis growth is
characterized by the time dependence of the coordinates $x_{0.2}$, $x_{0.5}$
, $x_{0.8}$ (in mm) and the corresponding temperatures $T_{0.2}$, $T_{0.5}$,
$T_{0.8}$ (in degrees Celsius). Different conditions representing various
possible limiting cases are considered. Namely, the tissue phantom without
response to temperature variations, with immediate and delayed response is
analyzed. In the first case the blood perfusion rate remains unchanged, $%
j=j_{v}=j_{0}.$ For tissue with immediate response ($\tau_{v}=0$) the
perfusion rate can attain large values directly at the beginning of the
necrosis growth, whereas for tissue with delayed response ($\tau_v>0$) this
increase will occur only after a lapse of time of order $\tau _{v}$.

\begin{figure}[tbp]
\par
\begin{center}
\includegraphics[width=85mm]{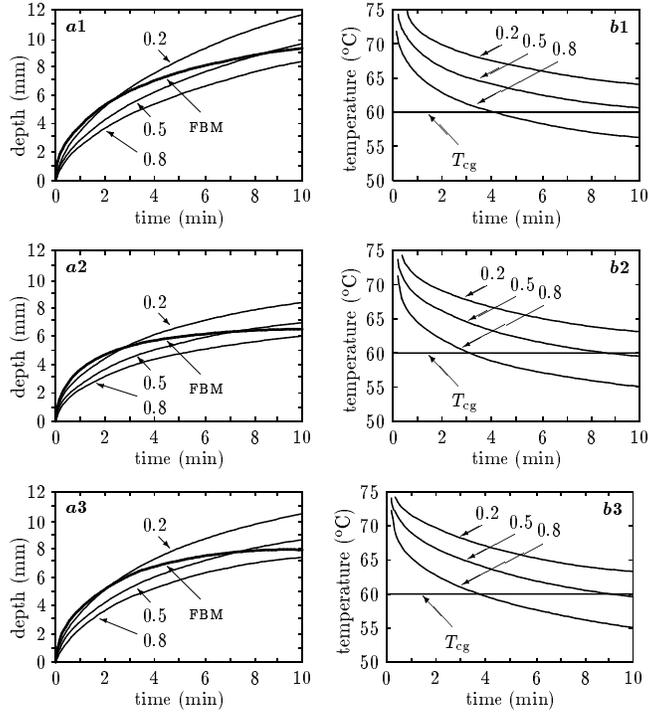}
\end{center}
\caption{The coordinates $x_{0.2}$, $x_{0.5}$, $x_{0.8}$ of the points at
which the undamaged tissue fraction $\protect\zeta =0.2,0.5,0.8$ ($a1$--$a3$%
) and the corresponding temperatures $T_{0.2}$, $T_{0.5}$, $T_{0.8}$ ($b1$--$%
b3$) as functions of time for different value of the parameters $\protect%
\alpha$, $\protect\tau_v$. In fig.~($a1$--$a3$) the thick lines labeled with
FBM are the position $\Re$ of the necrosis domain interface in the free
boundary model with the temperature coagulation $T_{\text{cg}}$ shown in
fig.~($b1$--$b3$). (For fig.~$a1$, $b1$ -- \{$\protect\alpha =1$\}; for fig.~%
$a2$, $b2$ -- \{$\protect\alpha = 0.3$, $\protect\tau_v = 0$\}; for fig.~$a3$%
, $b3$ -- \{$\protect\alpha = 0.3$, $\protect\tau_v = 3$~min\}; ($\protect%
\lambda_v =2$).}
\label{ThC:F9}
\end{figure}

Fig.~\ref{ThC:F9} shows the time dependence of the quantities $x_{0.2}$, $%
x_{0.5}$, $x_{0.8}$, and $T_{0.2}$, $T_{0.5}$, $T_{0.8}$ for the tissue
phantom without thermoregulation ($j_{\max }/j_{0}=1$), with immediate
strong response to temperature variations ($j_{\max }/j_{0}=5$, $\tau _{v}=0$%
) and for the tissue where the delay in the temperature response has a
pronounced effect on thermal coagulation ($\tau_v=2$~min). As seen in Fig.~%
\ref{ThC:F9}($b1$--$b3$) the decrease of the temperature $T_{0.5}$ becomes
slow sufficiently quickly (within approximately $1-2$~min) and the
temperature $T_{0.5}$ remains inside the interval $60-70~{}^{\circ}$C during
all the time of order of $5-10$~min which is a typical duration $t_{\text{tr}%
}$ of thermal treatment based on thermal coagulation. The width of this
interval is sixth as less as the typical value of the tissue overheating, $%
T_{b}-T_{a}\sim 60~^{\circ}$C. So at first approximation, the temperature $%
T_{0.5}$ may be treated as a fixed constant $T_{\text{cg}}$. Besides, whence
it follows that the coagulation temperature $T_{\text{cg}}$ being a
phenomenological parameter of this approximation can be estimated from the
expression
\begin{equation}
\frac{1}{\omega (T_{\text{cg}})}\sim t_{\text{tr}}\,.  \label{ThC:e3.19}
\end{equation}
Indeed, under such conditions thermal coagulation proceeds practically at
the temperature $T_{\text{cg}}$ and the value of $1/\omega (T_{\text{cg}})$
is approximately the time it takes for the live tissue located before in the
necrosis region to be damaged. Because of a strong temperature dependence of
$\omega (T)$ this estimate gives us the value of $T_{\text{cg}}$ to
sufficient accuracy. In addition, to make the comparison of the two model
more clear we have used in the simulation the value $T_{\text{cg}}=60$~$^{
\circ}$C found from expression~(\ref{ThC:e3.19}) (for $t_{\text{tr}}\sim 5 $%
~min) rather than the value of $T_{\text{cg}}$ approximating the dependence $%
T_{0.5}(t)$ to the best degree.

Another characteristics of thermal coagulation is illustrated in Fig.~\ref
{ThC:F9}($a1$--$a3$). Comparing the time dependence $x_{0.5}(t)$ with the
curves~``FBM'' (describing the motion $\Re(t)$ of the necrosis interface in
the free boundary model with the fixed coagulation temperature) we see that
there are two stages of the necrosis growth. The former corresponds to the
time interval $(0,t_{\text{cg}}),$ where $t_ {\text{cg}}\sim 1/j_{\text{max}%
}\sim 4$~min. At this stage the necrosis domain grows fast enough and in the
free boundary model the interface $\Gamma $ reaches its limit position $\Re
_{\text{lim}}$. This saturation of the interface displacement is due to the
temperature distribution becoming stationary. At the latter stage (from $t_{%
\text{cg}}$ to $t_{\text{tr}}$) the necrosis domain grows slowly and in the
free boundary model it is fixed. In other words, the proposed model makes
the difference of these stages more pronounced. If the treatment is
continued, the real necrosis domain will grow further and after a lapse of $%
20-30$~minutes the necrosis domain will deviate significantly in form from
that predicted by the given model. However such a prolonged treatment is
typically used to produce a hyperthermia effect (without visible injury)
rather than to cause thermal coagulation directly \cite{J94}.

It should be noted that the existence of the two stages does not obviously
result from the dependence $T_{0.5}(t)$ because in fact the necrosis
continues to grow slow at the second stage too. However, as follows from the
present analysis, these stages differ from each other not only in the
necrosis growth rate but also according to the behavior of the temperature
distribution. The slow stage is characterized by the quasistationary
temperature distribution. In other words, at this stage time scales on which
the size $\Re$ of the necrosis domain increases substantially are much
larger than time scales on which temperature distribution becomes
steady-state provided the necrosis boundary is fixed. The given property is
caused by the exponential dependence of the damaged tissue rate $\omega(T)$
on the temperature. At the fast stage these scales are of the same order.

Concluding this subsection we note that the fixed boundary temperature
approximation predicts the dynamics of thermal coagulation to a sufficiently
good accuracy. Indeed, the motion of the partially damaged tissue layer is
practically represented in Fig.~\ref{ThC:F9}($a1$--$a3$) by the region
bounded by the curves $x_{0.2}(t)$ and $x_{0.8}(t)$. So the free boundary
model can be regarded as adequate, if the difference $\left| x_{0.5}(t)-\Re
(t)\right| $ does not exceed the value $\left| x_{0.8}(t)-x_{0.2}(t)\right| $
remarkably. As seen in Fig.~\ref{ThC:F9}($a1$--$a3$) this is the case except
for the beginning of the growth and times larger than 10~min, where we have
gone beyond the fixed coagulation temperature approximation.

In the next section basing on the proposed model we will study the specific
properties of the necrosis growth limited by heat diffusion.

\section{Characteristics of the necrosis growth\label{ThC:sec.NG}}

The stated above model describes several different features of living tissue
affecting simultaneously the necrosis growth limited by heat diffusion. So
to penetrate deeper in its properties it is reasonable to analyze them
individually, which is the subject of the present section.

\subsection{Effect of the vessel discreteness\label{ThC:subsec.VD}}

To clarify the essence of this effect let us confine our consideration to
its qualitative analysis referring interested readers to specific papers~
\cite{we5,we7,we8} for a detailed analysis.

\begin{figure}[tbp]
\par
\begin{center}
\includegraphics[width=80mm]{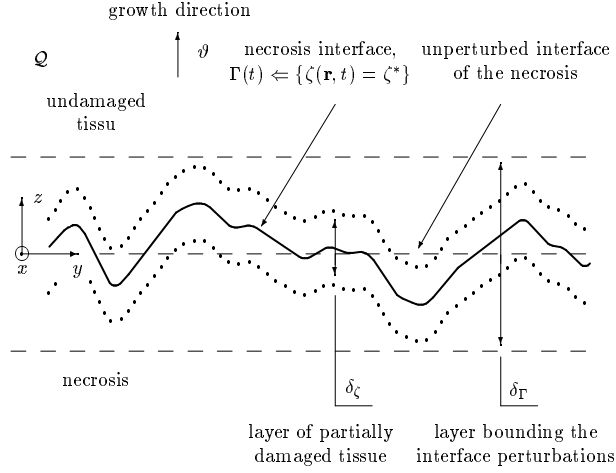}
\end{center}
\caption{Physical structure of the necrosis boundary. The necrosis interface
roughness caused by the vessel discreteness.}
\label{ThC:F10}
\end{figure}

The vessel discreteness manifesting itself in the random component $\delta j(%
\mathbf{r},t)$ of the blood perfusion rate causes random perturbation $%
\delta T(\mathbf{r},t)$ in the temperature field. Actually $\delta j$ is the
difference between the true perfusion rate and one averaged over scales
about $\ell _{v}$. So the latter is also the correlation length of field $%
\delta j(\mathbf{r},t)$ (Sec.~\ref{ThC:subsec.HM}). In this subsection we
mainly consider the formation of the necrosis domain of size $\Re \sim \ell
_{T}\sim \ell _{v}\sqrt{L_{n}}$, which enables us, at least formally, assume
the inequality $\Re \gg \ell _{v}$ to hold. The exception is the formation
of a three-dimensional necrosis domain whose growth is due to a small
applicator or the laser light absorption inside a small region~\cite{we4}
and it will be discussed below. Therefore in the case under consideration
the mean amplitude $\sigma $ and the correlation length $\lambda $ of the
temperature random nonuniformities $\delta T(\mathbf{r},t)$ should be
approximately the same as for such nonuniformities under uniform heating and
be given by formula~(\ref{ThC:e1.10}) and the estimate $\lambda \sim
\ell_{v} $. The field $\delta T(\mathbf{r},t)$ in its turn affects the form
of the necrosis interface $\Gamma$ making it randomly perturbed as shown in
Fig.~\ref{ThC:F10}. It is obviously, that the correlation length $\ell
_{\Gamma }$ of these interface perturbations is also about $\ell_v$, i.e.
\begin{equation}
\ell _{\Gamma }\sim \ell _{v}\sim \frac{1}{\sqrt{L_{n}}}\ell _{T}\sim \frac{%
1 }{\sqrt{L_{n}}}\Re \,.  \label{ThC:e4.0}
\end{equation}
The remaining characteristic of the interface perturbations that should be
found is their amplitude or, what is the same, the thickness $%
\delta_{\Gamma} $ of the layer bounding these perturbations (Fig.~\ref
{ThC:F10}).

In the fixed coagulation temperature approximation (Sec.~\ref{ThC:subsec.FT}%
) the form of the necrosis domain perturbed by the random temperature
nonuniformities $\delta T(\mathbf{r},t)$ due to the vessel discreteness is
specified by the condition:
\begin{equation}
\left. \left[ T_{v}(\mathbf{r},t)+\delta T(\mathbf{r},t)\right] \right| _{%
\mathbf{r}\in \Gamma }=T_{\text{cg}}\,,  \label{ThC:e4.1}
\end{equation}
where $T_{v}(\mathbf{r},t)$ is the regular component of the temperature
field (i.e. the solution of equations~(\ref{ThC:e3.4}), (\ref{ThC:e3.5}) for
$\delta j=0$). In a neighborhood of the necrosis boundary $\Gamma $ whose
thickness is much less than the mean necrosis radius $\Re$ the spatial
variations of the averaged temperature $T_{v}(\mathbf{r},t)$ can be
approximated by a linear dependence on the spatial coordinates $\mathbf{r}$:
\begin{equation}
T_{v}(\mathbf{r},t)\approx T_{\text{cg}}-G\left( r-\Re \right) \,.
\label{ThC:e4.2}
\end{equation}
Here $G\sim $ $(T_{\text{max}}-T_{a})/\Re \sim (T_{\text{cg}}-T_{a})/\ell
_{T}$ is the temperature gradient near the necrosis boundary $\Gamma $ and $%
r=\left| \mathbf{r}\right| $ provided the origin, $\mathbf{r}=0$, is placed
at the necrosis center. Then from (\ref{ThC:e4.1}) and (\ref{ThC:e4.2}) we
obtain $G\delta _{\Gamma }\sim \sigma $ whence taking also into account
expressions~(\ref{ThC:e1.7a}), (\ref{ThC:e1.10}) we get the following
estimate chain:
\begin{equation}
\delta _{\Gamma }\sim \frac{1}{\sqrt{L_{n}}}\ell _{v}\sim \frac{1}{L_{n}}
\ell _{T}\sim \frac{1}{L_{n}}\Re \,.  \label{ThC:e4.3}
\end{equation}
Expressions (\ref{ThC:e4.0}) and (\ref{ThC:e4.3}) substantiate, at least
formally, the validity of the following scale hierarchy: $\delta _{\Gamma
}<\ell _{\Gamma }<\Re $. So, when describing the growth of a necrosis domain
as a whole one can ignore the effect of the vessel discreteness and identify
the averaged tissue temperature $T_{v}(\mathbf{r},t)$ and the true one $T(%
\mathbf{r},t)$ \cite{we8}. However when the attention is focused on the
particular form of a necrosis domain, i.e. spatial scales smaller than the
mean necrosis size are under consideration such necrosis perturbations are
substantial.

To justify the latter statement and to measure the effect of these
perturbations on the necrosis form we compare the value $\delta _{\Gamma }$
with the thickness $\delta _{\zeta }$ of the layer of partially damaged
tissue where thermal coagulation is under way. As results from (\ref
{ThC:e1.17}) and (\ref{ThC:e4.3}) for typical values of the tissue
parameters we get
\begin{equation}
\frac{\delta _{\Gamma }}{\delta _{\zeta }}\sim \frac{(T_{\text{cg}}-T_{a})}{
L_{n}\Delta }\sim 2.  \label{ThC:e4.4}
\end{equation}
Therefore, these random perturbations of the necrosis interface exceed
sufficiently its physical thickness, i.e. the thickness of the layer of
partially damaged tissue. In other words, because of the temperature
nonuniformities due to the vessel discreteness the necrosis form should be
disturbed substantially, deviating remarkably from that predicted by the
distributed model dealing solely with the averaged tissue temperature $T_{v}(%
\mathbf{r},t)$.

However when, for example, the characteristic size of the applicator is
sufficiently small (about a few millimeters) the temperature field becomes
substantially nonuniform not only due to heat dissipation caused by blood
perfusion but also because of the geometric factor caused by the heat
propagation from the applicator into the surrounding tissue. In
three-dimensional space the temperature field $T(\mathbf{r})$ induced by a
small localized source even inside a tissue without perfusion will decrease
as $1/r$. Under such conditions it is this spatial decrease in temperature
that mainly controls the necrosis growth rather than heat diffusion into the
surrounding live tissue. So the thickness $\Re$ of the necrosis domain is
less then $\ell_{T}$, namely, can be about $\ell_{v}$\cite{we4} and the
perturbations of its boundary due to the vessel discreteness are depressed
considerably \cite{we7}.

Expressions (\ref{ThC:e4.0}) and (\ref{ThC:e4.3}) are actually the main
results of the present subsection. In the following subsections we will
focus out attention on the growth of a necrosis domain as a whole and, thus,
will ignore the vessel discreteness effect.

\subsection{Effects of the tissue response to heating\label{ThC:subsec.TR}}

In this section we analyze how the tissue response to local strong heating
affects the necrosis growth. To single out these effects in their own right
and to simplify the analysis we confine ourselves to studying the necrosis
growth in the one-dimensional tissue phantom described in Sec.~\ref
{ThC:subsec.MC} (Fig.~\ref{ThC:F5}) and adopt the fixed coagulation
temperature approximation.

\begin{figure}[tbp]
\par
\begin{center}
\includegraphics[width=85mm]{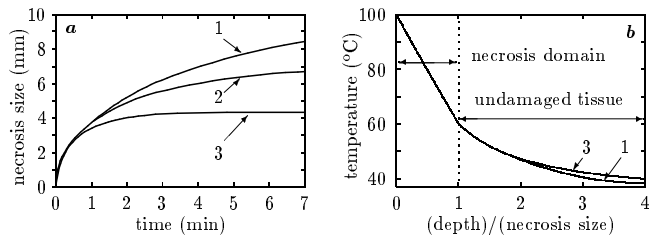}
\end{center}
\caption{The size $\Re $ of the necrosis domain vs time ($a$) and the
temperature distribution ($b$) for different values of $\protect\alpha $.
Curves~1,~2,~3 correspond to $\protect\alpha ={}$ 1, 0.3, 0.1, respectively (%
$\protect\tau _{v}=0$, $\protect\lambda _{v}=2$). }
\label{ThC:F11}
\end{figure}

First, we consider the tissue with immediate response, corresponding to $%
\tau_v=0$ in equation~(\ref{ThC:e3.2}). In this case the thickness $\Re_{%
\text{lim}}$ of the necrosis domain attained during the fast stage of the
coagulation is determined by the stationary temperature distribution. In
particular, for the tissue phantom without thermal regulation ($\alpha =1$)
the blood perfusion rate is constant, $j=j_{0}$, and as follows from
equations~(\ref{ThC:e3.4}) and (\ref{ThC:e3.5}):
\begin{equation}
\Re _{\text{lim}}^{\alpha =1}=\sqrt{\frac{F\kappa }{c\rho fj_{0}}}\cdot
\frac{T_{b}-T_{\text{cg}}}{F(T_{\text{cg}}-T_{a})}\sim 1\,\text{cm}\,.
\label{ThC:e5.1}
\end{equation}
For the tissue with thermal regulation expression~(\ref{ThC:e5.1}), after
the replacement $j_{0}\rightarrow j_{\text{max}}$, may be also used to
estimate the value $\Re_{\text{lim}}$, thus
\begin{equation}
\Re _{\text{lim}}\sim \sqrt{\frac{F\kappa }{c_{t}\rho _{t}fj_{\text{max}}}}
\,.  \label{ThC:e5.2}
\end{equation}
The dynamics of coagulation under such conditions is represented in Fig.~\ref
{ThC:F11} for different values of the parameter $\alpha=j_{0}/j_{\text{max}}$%
.

Fig.~\ref{ThC:F11}$a$ shows the size $\Re$ of the necrosis domain as
function of time. The higher is the tissue response, the smaller is the
necrosis domain and the shorter is the fast stage of the necrosis growth.
The duration of the fast stage is actually the characteristic time $t_{\text{%
cg}}$ during which the size of the necrosis domain attains values about $\Re
_{\text{lim}}$. Comparing the numerical values of the corresponding
quantities we find that the duration of the fast stage can be estimated by
the expression
\begin{equation}
t_{\text{cg}}\sim \frac{1}{fj_{\text{max}}}\,,  \label{ThC:e5.3}
\end{equation}
which conforms to the general properties of heat transfer in living tissue.
Indeed, in the given model, as follows from equations~(\ref{ThC:e3.4}) and (%
\ref{ThC:e3.5}), the time it takes for the temperature distribution to
become steady-state is about $fj_{\text{max}}$ and the establishment of this
steady-state (in reality, quasistationary) temperature distribution is
actually the essence of the fast stage.

Fig.~\ref{ThC:F11}$b$ shows the temperature distribution for the tissue
without thermoregulation (curve~1) and for the tissue with strong immediate
response (curve 2, $j_{\max }/j_{0}=10$). In order to compare them with each
other, lengths are measured in units of the corresponding necrosis size.
Whence it follows that, in contract to the time dependence $\Re (t)$, the
tissue response to heating practically does not affect the form of
temperature distribution.

\begin{figure}[tbp]
\par
\begin{center}
\includegraphics[width=85mm]{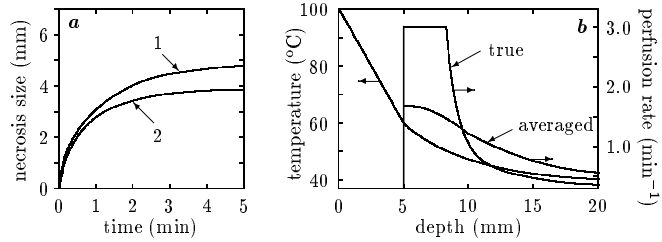}
\end{center}
\caption{($a$) The size $\Re$ of the necrosis domain as the function of time
described by the stated free boundary model (curve~1) and by the same model
within the replacement $j_v \rightarrow j$ (curve~2). ($b$)~The distribution
of the temperature $T$, the true perfusion rate $j$, and the averaged one $%
j_v$ corresponding to curve~1 at time $t=2$~min. ($\protect\alpha=0.1$, $%
\protect\tau_v = 0$, $\protect\lambda_v = 2$, $T_{\text{vr}}=50~{}^{\circ}$%
C) }
\label{ThC:F12}
\end{figure}

When the tissue response to temperature variations is sufficiently strong ($%
\alpha \ll 1$) the blood perfusion rate $j$ becomes substantially
nonuniform. In this case the averaged perfusion rate $j_{v}$ differs
significantly from the true one $j$, which has a certain effect on the
growth of small necroses. The latter feature is illustrated in Fig.~\ref
{ThC:F12}

Fig.~\ref{ThC:F12}$a$ compares the dynamics of the necrosis growth described
by the given model and by the same model where, however, equation~(\ref
{ThC:e3.9}) is omitted and the replacement $j_{v}\rightarrow j$ is made.
Fig.~\ref{ThC:F12}$b$ shows the distribution of the tissue temperature $T$,
the true blood perfusion rate $j$, and the averaged one $j_{v}$ that occur
when the tissue responds in such an intensive way. In this case, as seen in
Fig.~\ref{ThC:F12}$b$, the averaged blood perfusion rate can be twice as
less as the true one. The latter has a certain effect on the necrosis growth
because ignoring the difference between $j_{v}$ and $j $ gives a more lower
estimate of the necrosis size (Fig.~\ref{ThC:F12}$a$).

\begin{figure}[tbp]
\par
\begin{center}
\includegraphics[width=85mm]{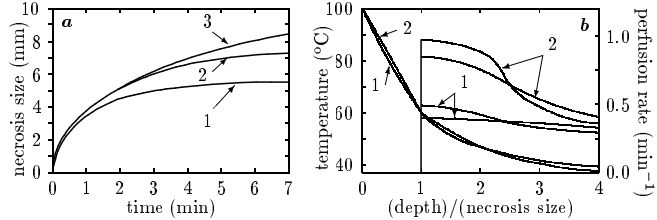}
\end{center}
\caption{($a$) The size $\Re $ of the necrosis domain vs time for the tissue
responding immediately (curve~1) and with a certain delay (curve~2, $\protect%
\tau _{v}=2$~min). Curve~3 represents the $\Re (t)$-dependence for the
tissue without thermal regulation. ($b$) The distribution of the
temperature, true and averaged perfusion rates at different time moments $%
t=1 $~min (curves~1) and $t=3$~min (curves~2). ($\protect\alpha =0.2$, $%
\protect\lambda _{v}=1$)}
\label{ThC:F13}
\end{figure}

\begin{figure}[tbp]
\par
\begin{center}
\includegraphics[width=60mm]{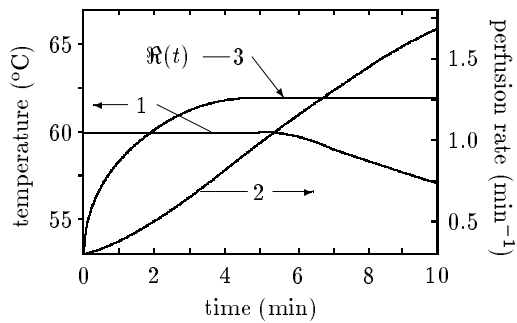}
\end{center}
\caption{The temperature (curve~1) and the averaged perfusion rate (curve~2)
at the necrosis interface $\Gamma$ vs time for the tissue with the delayed
response. Curve~3 shows the corresponding time dependence of the necrosis
size $\Re (t)$. ($\protect\tau_v = 1.4$~min, $\protect\alpha = 0.07$, $%
\protect\lambda_v=0.5$)}
\label{ThC:F14}
\end{figure}

Now we consider how the delay in the tissue response can affect thermal
coagulation. This effect is remarkable when the delay time $\tau _{v}$ is
comparable with the duration of the fast stage $t_{\text{cg}}$. So we may
confine ourselves to values $\tau _{v}\sim $ $t_{\text{cg}}$. The difference
in dynamics of the necrosis domain growth for the tissues responding
immediately ($\tau _{v}=0$) and with a certain delay ($\tau _{v}\sim 2$~min)
is illustrated in Fig.~\ref{ThC:F13} and Fig.~\ref{ThC:F14}.

As seen in Fig.~\ref{ThC:F13}$a$ when the tissue response is delayed
essentially the necrosis domain, firstly, grows fast enough keeping ahead of
one growing in the tissue with the immediate response. Then the growth of
the necrosis domain is suppressed and further its form remains unchanged. It
should be noted that in this case the saturation of the necrosis domain
growth is not due to the temperature distribution becoming stationary. Its
reason is the increase of the blood perfusion rate after a lapse of a time
about $\tau $. Under such conditions there is enough time for the size $\Re $
of the necrosis domain to attain values of order $\Re _{\text{lim}}^{\alpha
=1}\propto \sqrt{1/j_{0}}$ until the blood perfusion rate increases
substantially. These values could not be attained if the tissue response
were not delayed. So after the blood perfusion rate increases, the following
growth of the necrosis domain becomes impossible and the temperature $T_{i}$
at the interface $\Gamma $ must go below $T_{\text{cg}}$ and the necrosis
domain has to cease to grow. This behavior of the interface temperature is
illustrated in Fig.~\ref{ThC:F14}. It should be noted that such a saturation
of the necrosis growth for real tissues seems to be more pronounced because
a real necrosis continues to grow at the slow stage until the blood
perfusion rate becomes high enough.

Fig.~\ref{ThC:F13}$b$ shows the distribution of the temperature and blood
perfusion rate at different time moments for the tissue with delayed
response. As before, the length is measured in the corresponding values of
the necrosis size in order to compare these distributions. As time elapses
the perfusion rate increases due to the tissue response. In contrast to this
behavior of the perfusion rate, the form of the temperature distribution
practically remains unchanged. The comparison of the given result and one
obtained for the tissue responding to temperature variations immediately
(Fig.~\ref{ThC:F11}$b$) leads us to the conclusion that the form of the
temperature distribution occurring in tissue during the necrosis growth
depends weakly on the specific values of the tissue parameters. This
conclusion, in particular, forms the basis for applying variational
techniques to analysis of the thermal coagulation due to laser-tissue
interaction.

It should be noted that the conclusion concerning the universal form of the
temperature distribution has been made analyzing the necrosis growth in the
tissue phantoms that differ only in the properties of thermoregulation. The
other tissue parameters (for example, the tissue thermal conductivity $%
\kappa $ and the initial value $j_{0}$ of the blood perfusion rate) take on
particular values fixed in the present analysis. This raises the question of
whether the stated conclusion will hold if we change these parameters too.
However, choosing the appropriate units of time and length aggregating such
parameters we can rewrite the governing equations in the dimensionless form.
Thus their particular values are not the factor.

\subsection{Effect of the applicator geometry}

In the previous section we have actually studied the characteristics of the
necrosis growth when the applicator is sufficiently large and the necrosis
domain can be locally treated as a plane layer. If the applicator is small
enough so its size is about or even less than the thickness of the necrosis
layer the heat diffusion into the surrounding tissue gives rise to the
temperature distribution depending on the applicator form. The latter in
turn affects the necrosis formation. This effect is the subject of the
present section reviewing the results obtained in \cite{we4}.

As in the previous section we study the necrosis growth in the tissue
phantom described in Sec.~\ref{ThC:subsec.MC} (Fig.~\ref{ThC:F5}) but
applying the free boundary model in its general form (Sec.~\ref{11}), i.e.
assuming the necrosis boundary motion to be governed by expression~(\ref
{ThC:e3.7}) because under such conditions time variations in the coagulation
temperature become remarkable. However, keeping in mind that the latter fact
is the main reason of the effects under consideration we will make no
difference between the true and averaged perfusion rates in order to
simplify the analysis.

\begin{figure}[tbp]
\par
\begin{center}
\includegraphics[width=85mm]{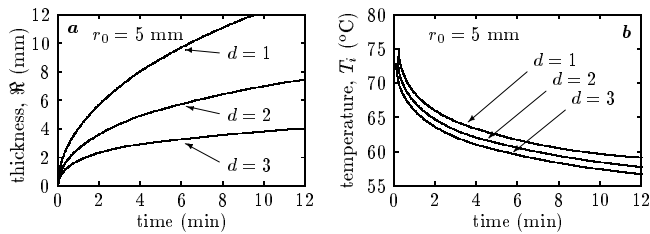}
\end{center}
\caption{Thickness $\Re (t)$ of the necrosis layer ($a$) and temperature $%
T_{i}(t)$ at the necrosis interface ($b$) vs time $t$ for heat sources of
the plane ($d=1$), cylindrical ($d=2$), and spherical ($d=3$) form. Here we
consider the tissue without response and $r_{0}=5$~mm.}
\label{ThC:F15}
\end{figure}

\begin{figure}[tbp]
\par
\begin{center}
\includegraphics[width=85mm]{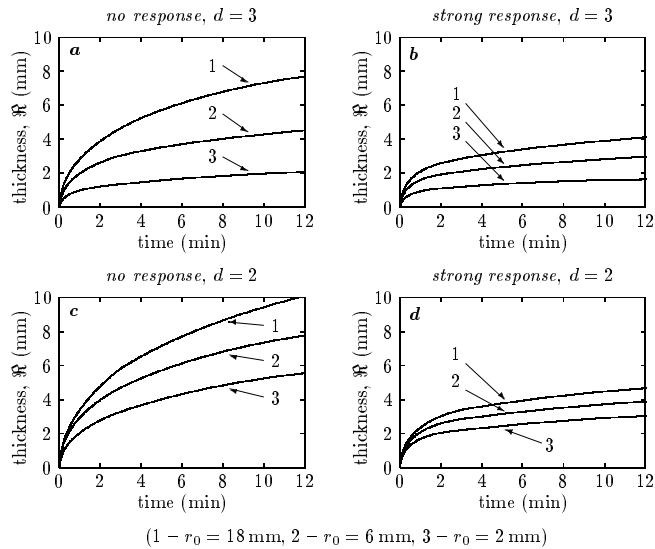}
\end{center}
\caption{Thickness $\Re (t)$ of the spherical ($a$, $b$) and cylindrical ($c$%
, $d$) necrosis layers vs time $t$ for different values of the initial
necrosis radius $r_{0}=18$~mm (curve 1), $r_{0}=6$~mm (curve 2), and $%
r_{0}=2 $~mm (curve 3) ($a$,~$c$~--~tissue without response, $b$,~$d$%
~--~tissue with strong ($j_{\text{max}}=10j_{0}$) immediate response).}
\label{ThC:F16}
\end{figure}

\begin{figure}[tbp]
\par
\begin{center}
\includegraphics[width=85mm]{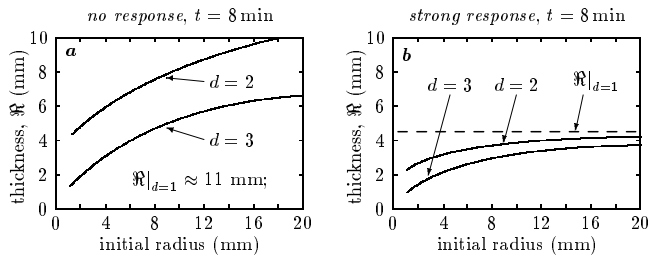}
\end{center}
\caption{Thickness $\Re$ of the necrosis layer at the fixed moment of time $%
t = 8$~min for different values of the initial necrosis radius $r_0$ for the
two- and three-dimensional tissue phantom. ($a$~--~tissue without response, $%
b$~--~tissue with a strong ($j_{\text{max}}=10j_0$) immediate response)}
\label{ThC:F18}
\end{figure}

Characteristic features of the necrosis growth depending actually on the
applicator form are demonstrated in Fig.~\ref{ThC:F15}--Fig.~\ref{ThC:F18}.
Namely, Fig.~\ref{ThC:F15}$a$ shows how the time dependence of the necrosis
layer thickness $\Re (t)$ changes for the applicator of plane ($d=1$),
cylindrical ($d=2$), and spherical ($d=3$) form for the tissue phantom with
the same properties. Here is illustrated the dynamics of the necrosis growth
in the tissue phantom without response to temperature variations for $%
r_{0}=5 $~mm. Fig.~\ref{ThC:F15}$a$ shows that for one-dimensional tissue
phantom the difference between the fast and the slow stages is not well
distinctive and the necrosis growth is under way practically throughout the
whole course of a typical thermotherapy procedure. For the three-dimensional
tissue phantom we meet the opposite situation. In this case the necrosis
growth exhibits the clearly evident tendency to saturation after a lapse of
several minutes and after 12 minutes the thickness $\left. \Re \right|
_{d=3} $ of the necrosis layer can attain the value only somewhat less than
the initial radius $r_{0}$ of the necrosis domain. At the same time in the
one-dimensional tissue phantom the necrosis radius exceeds this value by
several times. In the two-dimensional tissue phantom the necrosis growth is
characterized by the intermediate behavior. It should be noted that in
contrast to this the time dependence of the interface temperature $T_{i}(t)$
(Fig.~\ref{ThC:F15}$b$) is practically of the same form for the three cases.

These results prompt us that the rate of the necrosis growth in the
three-dimensional tissue phantom should depend substantially on the initial
necrosis radius $r_{0}$ at least for $r_{0}\ll \left. \Re \right| _{d=1}$,
where $\left. \Re \right| _{d=1}$ is the thickness that the necrosis layer
would attain for an applicator of plane geometry. This fact is directly
demonstrated in Fig.~\ref{ThC:F16} showing the dynamics of the necrosis
growth in the tissue phantom without response to temperature variations
(figure~$a$) as well as with an immediate strong response (figure~$b$). At a
fixed moment of time the necrosis layers can differ in thickness by a factor
of three to four as the initial radius $r_{0} $ changes from 2~mm to 18~mm.
As should be expected for the two-dimensional tissue phantom this dependence
is smoothed (Fig.~\ref{ThC:F16}) and under the same conditions the necrosis
thickness can increase not more than two-fold.

Fig.~\ref{ThC:F18} illustrates the dependence of the necrosis growth on the
initial necrosis radius in the most clear form. These curves have been
obtained by fixing the time $t=8$~min and treating the point collection $%
\left\{ \left. \Re \right| _{t=8~\text{min}},r_{0}\right\} $ for different
values of $r_{0}$ as the partition points of certain continuous curves. In
other words, this figure shows the thickness $\left. \Re (t,r_{0})\right|
_{t=8~\text{min}}$ of the necrosis layer as a function of its initial radius
$r_{0}$ for the fixed moment of time under various physical conditions. We
see in Fig.~\ref{ThC:F18} that for the three-dimensional tissue phantom the
curve $\left. \Re \right| _{d=3}(r_{0})$ practically goes into the origin of
the plane $\{\Re ,r_{0}\}$ as formally $r_{0}\rightarrow 0$. When the
initial radius $r_{0}$ becomes large enough ($r_{0}>\left. \Re \right|
_{d=1} $) the curve~$\left. \Re \right| _{d=3}(r_{0})$ (as it must) tends to
the value $\left. \Re \right| _{d=1}$ (figure~{$b$). In other words, until $%
r_{0}<\left. \Re \right| _{d=1}$ it is the initial radius of the necrosis
domain that directly controls the size of the necrosis layer which can be
attained during a typical course of thermotherapy. This property of the
necrosis growth is due to the fact that in the three-dimensional space the
temperature field governed by heat diffusion from a local source would
remain substantially nonuniform ($T(r)\propto \frac{1}{r}$ as $r\rightarrow
\infty $) even though we ignored the heat sink effect caused by blood
perfusion. So in the three-dimensional tissue phantom the temperature at the
necrosis interface $T_{\text{cg}}$ inevitably has to decrease substantially
as the necrosis layer grows. So due to the strong temperature dependence of
the coagulation rate $\omega (T_{\text{cg}})$ the necrosis growth will be
practically suppressed after a lapse of time it takes for the $\left. \Re
\right| _{d=1}(t)$ to exceed the initial necrosis radius $r_{0}$. Whence it
follows that for real applicators of the spherical form whose size does not
exceed several millimeters the thickness of the necrosis layer will be
directly controlled by its size and, may be, by the depth of laser light
penetration into the tissue because in this case heat diffusion standing
along leads to the saturation in the necrosis growth. In the two-dimensional
space (and more so in the one-dimensional space) heat diffusion tends to
make the temperature field uniform. Thus, for the two-dimensional tissue
phantom the dependence of the necrosis growth on the initial radius should
be smoothed and the thickness of the necrosis layer will be finite even
though we formally set $r_{0}=0$ (Fig.~\ref{ThC:F18}). In particular, for
the case shown in Fig.~\ref{ThC:F18} the change of the initial radius from
zero to infinity causes the necrosis layer in the two-dimensional tissue
phantom to increase in thickness by 2--3-fold. }

\begin{figure}[tbp]
\par
\begin{center}
\includegraphics[width=60mm]{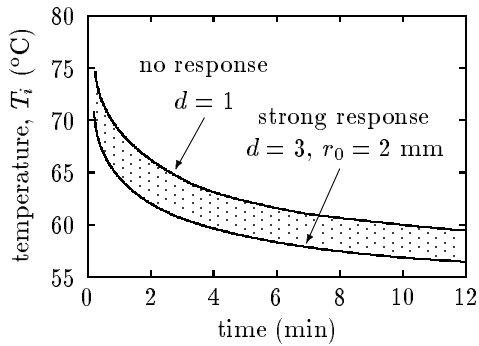}
\end{center}
\caption{The universality of the time dependence of the coagulation
(interface) temperature $T_{i}(t)$. All the curves $T_{i}(t)$ obtained under
various conditions belong to the dotted region. The upper boundary of this
region corresponds to the one-dimensional tissue phantom without temperature
response, the lower one corresponds to the tree-dimensional tissue tissue
phantom with a strong immediate response ($j_{\text{max}}=10j_{0}$).}
\label{ThC:F19}
\end{figure}

The dynamics of the necrosis growth depends substantially on the physical
conditions. However, it turns out that the time dependence of the
coagulation temperature $T_{\text{cg}}(t)$ (i.e. the temperature at the
necrosis interface) is practically insensitive to particular details of the
growth conditions. This fact is demonstrated in Fig.~\ref{ThC:F19}. We have
discovered that all the curves $T_{\text{cg}}(t)$ obtained numerically for
various growth conditions are located inside the dotted region which is thin
enough. This property of the necrosis growth is relative to the universality
of the one-dimensional temperature distribution demonstrated in Sec.~\ref
{ThC:subsec.TR} and justify the conclusion that the general form of the
temperature field can depend slowly on particular values of the main tissue
parameters. Such a fact shows the feasibility of applying variational
methods to analysis of the necrosis growth in more complicated cases and is
undoubtedly worth special consideration.

\subsection{Self-localization of the necrosis growth in tissue with a tumor
due to thermoregulation}

In section~\ref{ThC:subsec.TR} we have demonstrated that the tissue response
to heating affects substantially the necrosis growth limited by heat
diffusion. It is due to the caused increase in the blood perfusion rate
strengthening the heat dissipation in the surrounding undamaged tissue,
which suppresses the further heat diffusion into it and, as the result, the
necrosis growth. According to the experimental data \cite{Song84} only the
normal tissue can exhibit a significant increase in the blood perfusion rate
in the response to the temperature growth. whereas inside a tumor an
increase in the blood perfusion rate is depressed. This leads to some
self-localization of the temperature inside tumor and self-localization of
necrosis growth due to thermoregulation. The latter feature is illustrated
in Fig.~\ref{b1}, \ref{b2} where results of mathematical modeling of the two
and three dimensional cases are presented.

\FRAME{ftbpFU}{3.013in}{3.1609in}{0pt}{\Qcb{The stationary distribution of
the temperature in living tissue containing necrosis domain of the size $\Re
=0.5$ for two-dimensional case: ($a$) $q/c\protect\rho =10\exp (-10y)$,  ($b$%
)- $q/c\protect\rho =10\exp (-5r),r=\protect\sqrt{x^{2}+y^{2},}\protect%
\alpha =0.05,$}}{\Qlb{b1}}{B1.JPG}{\special{language "Scientific Word";type
"GRAPHIC";maintain-aspect-ratio TRUE;display "USEDEF";valid_file "F";width
3.013in;height 3.1609in;depth 0pt;original-width 5.9067in;original-height
6.1981in;cropleft "0";croptop "1";cropright "1";cropbottom "0";filename
'B1.JPG';file-properties "XNPEU";}}


\FRAME{ftbpFU}{3.013in}{3.1609in}{0pt}{\Qcb{The stationary distribution of
the temperature in living tissue containing necrosis domain of the size $\Re
=0.5$ for three-dimensional case: ($a$)- $q/c\protect\rho =12\exp (-5r),r=%
\protect\sqrt{z^{2}+y^{2}}$ ($b$)-$q/c\protect\rho =20\exp (-5\protect\sqrt{%
r^{2}+x^{2}}),\protect\alpha =0.05.$}}{\Qlb{b2}}{B2.JPG}{\special{language
"Scientific Word";type "GRAPHIC";maintain-aspect-ratio TRUE;display
"USEDEF";valid_file "F";width 3.013in;height 3.1609in;depth
0pt;original-width 5.9067in;original-height 6.1981in;cropleft "0";croptop
"1";cropright "1";cropbottom "0";filename 'B2.JPG';file-properties "XNPEU";}}%
\bigskip


\section{Laser induced heat diffusion limited tissue coagulation as a
specific therapy mode}

One of the main parameters controlling a thermotherapy treatment is the
power of irradiation delivered into the tissue and the treatment duration.
These parameters as well as a specific therapy mode should be chosen in such
a way that gives rise to the necrosis of desired form. Different modes, in
principle, can give rise to a necrosis region of the same form and, in this
case, a specific mode may be chosen keeping in mind, for example, its
optimality and stability. Various thermotherapy modes are singled out by
those physical mechanisms that play the leading role and endow them with
individual properties \cite{J94}. The presented results enable us to regard
the laser induced heat diffusion limited tissue coagulation as an individual
therapy mode where heat diffusion into the live tissue, i.e. the tissue with
active blood perfusion, plays the governing role.

To justify the latter statement let us briefly summarize the aforesaid.
Namely, we have considered thermal tissue coagulation induced by local
heating due to laser light absorption and limited by heat diffusion into the
surrounding live tissue. The irradiation power is assumed to be enough high
for the temperature inside the region affected directly by laser light to
grow up to values about 100$\,^{\circ}$C. Then this process is characterized
by the following. \vspace{0.3\baselineskip}

(I) In the considered case the necrosis growth is governed by the tissue
thermal coagulation inside a layer of thickness about 1~mm that separates
the necrosis domain and the live tissue and involves the partially damaged
tissue. This layer is thin enough so the standard bioheat equation cannot be
applied to describing temperature distribution in it because on such scales
this equation does not hold. Therefore, the approach used previously by
other authors (see, e.g., \cite{JM95,MR95} and references therein) which is
based on the continuous description of heat transfer and thermal coagulation
inside the layer of partially damaged tissue (the distributed model) is
inconsistent. \vspace{0.3\baselineskip}

(II) This inconsistency is overcome in the free boundary model of local
thermal coagulation which regards the layer of partially damaged tissue as
an infinitely thin interface of the necrosis domain whose motion is governed
by the boundaries values of tissue temperature and its gradient. In
addition, the free boundary model takes into account the perturbations of
the tissue temperature caused by the vessel discreteness. \vspace{%
0.3\baselineskip}

(III) The necrosis growth caused by heat diffusion limited thermal
coagulation depends weakly on the particular details of heat transfer inside
the layer of partially damaged tissue. This property justifies the
distributed model. Nevertheless, we think that using the free boundary model
is favored because in addition to being consistent it singles out the
characteristic features governing the necrosis growth. \vspace{%
0.3\baselineskip} 

In particular, basing on the free boundary model we have found that: \vspace{%
0.3\baselineskip}

(IV) Heat diffusion limited thermal coagulation involves two stages, fast
and slow. It is the former stage during which the necrosis domain is mainly
formed and its size $\Re $ can be estimated by expression~(\ref{ThC:e5.2}).
Expression~(\ref{ThC:e5.3}) estimates the duration $t_{\text{cg} }$ of this
stage. The latter stage is characterized by a substantially slower growth of
the necrosis domain, with this slowdown becoming more pronounced as the
applicator dimensions decrease. For the applicators of the spherical form
whose radius does not exceed several millimeters the characteristic size of
a necrosis domain formed during a typical course of thermotherapy is
directly determined by the applicator radius and, may be, by the depth of
laser light penetration into the tissue. Under such conditions even heat
diffusion itself gives rise to the practical saturation in the necrosis
growth and the fast and slow stages are most distinctive. \vspace{%
0.3\baselineskip}

(V) The blood perfusion rate can become extremely nonuniform in the vicinity
of the necrosis domain because of the strong tissue response to temperature
variations. In this case in modelling thermal coagulation one should take
into account that the temperature distribution is governed by the averaged
blood perfusion rate rather than the true one. The delay in the tissue
response on time scales of order $t_{\text{cg}}$ can affect thermal
coagulation essentially. In this case, in particular, the duration of the
fast stage is controlled by the delay time $\tau _{v}$. The size $\Re $
attained by the necrosis domain under such conditions is determined by the
initial value of the blood perfusion rate $j_{0}$ rather than those ($j_{%
\text{max}}\gg j_{0}$) occurring near the necrosis domain at the second
stage due to the temperature tissue response. \vspace{0.3\baselineskip}

(VI) The discreteness of the vessel arrangement causes perturbations in the
tissue temperature, which is described by introducing additional
nonuniformities of the blood perfusion rate regarded as random. These
temperature nonuniformities it turn give rise to perturbations in the
necrosis boundary whose amplitude $\delta _{\Gamma }$ and the correlation
length $\ell _{\Gamma }$ are estimated as:
\begin{equation*}
\delta _{\Gamma }\sim \frac{1}{L_{n}}\Re \,,\quad \ell _{\Gamma }\sim \frac{%
1 }{\sqrt{L_{n}}}\Re \,.
\end{equation*}
Here the factor $L_{n}\approx \ln (l/a)\approx 4$ ($l/a\sim 40$ is the
characteristic ratio of the individual length to radius of blood vessels
forming peripheral circulation systems). It should be noted that, first,
these boundary perturbations are remarkable because they exceed in amplitude
the thickness $\delta _{\zeta }$ of the layer of partially damaged tissue
where thermal coagulation is under way. Second, the universal form of these
relations is due to the vascular network being fractal in structure.
However, for spherical applicators of small size (about several millimeters)
the effect of the vessel discreteness is depressed because in this case
blood perfusion does not affect the necrosis growth substantially. \vspace{%
0.3\baselineskip} 

Concerning certain additional aspects in the theoretical description of heat
diffusion limited thermal coagulation we can state the following: \vspace{%
0.3\baselineskip}

(VII) The free boundary model can be the basis of a faster numerical
algorithm of simulating the necrosis growth because it deals with solely the
necrosis region and the undamaged tissue where the temperature distribution $%
T(\mathbf{r},t)$ is a smooth field. \vspace{0.3\baselineskip}

(VIII) The time dependence of the temperature in the layer where thermal
coagulation is under way (the coagulation temperature) is practically
insensitive to the particular details of the growth conditions and to the
particular values of the tissue parameters. For applicators of large
dimensions (exceeding 1\thinspace cm) the form of the temperature
distribution occurring in tissue during the necrosis growth also depends
weakly on the particular values of the tissue parameters. The given
properties form the basis for applying different variational techniques to
analyzing heat diffusion limited thermal coagulation. 

\section{Acknowledgments}

This work was supported by STCU grant \ \#1675.

\end{document}